\documentclass[fleqn,usenatbib]{mnras}

\usepackage{newtxtext,newtxmath}

\usepackage[T1]{fontenc}
\usepackage{ae,aecompl}
\usepackage{url}

\usepackage{graphicx}	
\usepackage{amsmath}	
\usepackage{pifont}
\usepackage{float}
\usepackage{color}

\newcommand {\magg}{\hphantom{>}}  
\newcommand{\xmark}{\ding{55}}%

\newcommand{\nup}{$\nu_{\rm peak}^S$}

\usepackage{lineno}
\usepackage{ulem}

\definecolor{darkspringgreen}{rgb}{0.09, 0.45, 0.27}

\title[The Spectra of IceCube Neutrinos II]{The Spectra of IceCube Neutrino (SIN) candidate sources  - II. Source Characterisation}

\author[P. Padovani et al.]{P. Padovani$^{1,2}$\thanks{E-mail:
ppadovan@eso.org}, 
P. Giommi$^{3,4,5}$, R. Falomo$^{6}$, F. Oikonomou$^7$, M. Petropoulou$^{8}$\thanks{Mercator Fellow}, T. Glauch$^9$,
\newauthor E. Resconi$^9$, A. Treves$^{10,11}$, S. Paiano$^{12,13,14}$\\
$^{1}$European Southern Observatory, Karl-Schwarzschild-Str. 
2, D-85748 Garching bei M\"unchen, Germany\\
$^{2}$Associated to INAF - Osservatorio di Astrofisica e Scienza dello Spazio, via Piero 
Gobetti 93/3, I-40129 Bologna, Italy\\
$^{3}$Associated to Agenzia Spaziale Italiana, ASI, via del Politecnico s.n.c., I-00133 Roma,  
Italy\\
$^{4}$Institute for Advanced Study, Technische Universit{\"a}t M{\"u}nchen,
Lichtenbergstrasse 2a, D-85748 Garching bei M\"unchen, Germany\\
$^{5}$Center for Astro, Particle and Planetary Physics, New York University, Abu Dhabi\\
$^{6}$INAF - Osservatorio Astronomico di Padova, vicolo dell'Osservatorio 5, I-35122, Padova, 
Italy\\
$^7$Department of Physics, NTNU, NO-7491 Trondheim, Norway \\
$^{8}$Department of Physics, National \& Kapodistrian University of Athens, GR-15784, Athens, Greece \\
$^{9}$Technische Universit{\"a}t M{\"u}nchen, Physik-Department, 
James-Frank-Str. 1, D-85748 Garching bei M{\"u}nchen, Germany\\
$^{10}$Universit\`a dell'Insubria, via Valeggio, 22100, Como, Italy\\
$^{11}$INAF - Osservatorio Astronomico di Brera, via Bianchi 46, I-23807, Merate (Lecco), Italy\\
$^{12}$INAF - Osservatorio Astronomico di Roma, via Frascati 33, I-00040, Monteporzio Catone, 
Italy \\
$^{13}$INAF - IASF Milano, via Corti 12, I-20133, Milano, Italy \\
$^{14}$INAF - IASF Palermo, via Ugo La Malfa, 153, I-90146, Palermo, Italy \\ 
}

\date{Accepted XX. Received YY; in original form ZZ}

\pubyear{2021}

\begin{document}
\label{firstpage}
\pagerange{\pageref{firstpage}--\pageref{lastpage}}
\maketitle

\begin{abstract}
Eight years after the first detection of high-energy astrophysical neutrinos 
by IceCube we are still almost clueless as regards to their origin, although 
the case for blazars being neutrino sources is getting stronger. 
After the first significant association at the $3 - 3.5\,\sigma$ level in time 
and space with IceCube neutrinos, i.e. the blazar TXS\,0506+056 at $z=0.3365$,
some of us have in fact selected a unique sample of 47 blazars, out of which 
$ \sim 16$ could be associated with individual neutrino track events 
detected by IceCube. Building upon our recent spectroscopy 
work on these objects, here we characterise them to determine their 
real nature and check if they are different from the rest of the blazar 
population. For the first time we also present a systematic study 
of the frequency of masquerading BL Lacs, i.e. flat-spectrum radio quasars 
with their broad lines swamped by non-thermal jet emission, in a $\gamma$-ray- 
and IceCube-selected sample, finding a fraction $>$ 24 per cent and 
possibly as high as 80 per cent. 
In terms of their broad-band properties, our sources appear to be 
indistinguishable from the rest of the blazar population. We also discuss 
two theoretical scenarios for neutrino emission, one in which neutrinos 
are produced in interactions of protons with jet photons and one in 
which the target photons are from the broad line region. Both scenarios can equally
account for the neutrino-blazar correlation observed by some of us. 
Future observations with neutrino telescopes and X-ray satellites will test them out.

\end{abstract}


\section{Introduction}\label{sec:Introduction}

The observation of ultra-high energy cosmic rays (CRs) by the Pierre Auger
Observatory (PAO) and Telescope Array (TA) has revealed the existence of
extreme cosmic accelerators but not yet their nature and location in the
Universe \citep[see, e.g.][and references therein]{Anchordoqui_2019}.
Complementary to the PAO and TA observations the IceCube Neutrino
Observatory\footnote{\url{http://icecube.wisc.edu}} at the South Pole has
detected tens of neutrinos of likely astrophysical and extragalactic origin
with energies extending beyond 1 PeV ($10^{15}$ eV) \citep[e.g.][and
  references therein]{schneider2019,stettner2019,Aartsen2020a}. Neutrinos of
such high energies are most likely generated by the interaction of very
high-energy (VHE) CRs with matter or radiation, which leads to the
production of charged and neutral mesons, which then decay into neutrinos,
$\gamma$-rays, and other particles. This has at least two important
astronomical implications: (1) at variance with CRs, which, being charged,
get deflected, and $\gamma$-rays, which are absorbed by pair-production
interactions with the extragalactic background light (EBL) at $E \gtrsim
100$ GeV (\citealt{Gould_1967}; see also, e.g. \citealt{biteau2020} 
and references therein), neutrinos
can travel cosmological distances basically unaffected by matter 
and magnetic fields and are the only
``messengers'', which can provide information on the VHE physical processes
that generated them. Said differently, the extragalactic photon sky is
almost completely dark at the energies sampled by IceCube ($\gtrsim 60$
TeV); (2) the presence of PeV neutrinos implies the existence of protons up
to energies $\gtrsim 10^{17}$\,eV. This has huge implications for the
study of high-energy emission processes in astronomical sources.

Various astrophysical classes have been suggested to be responsible
for the observed IceCube signal \citep[e.g.][and references
  therein]{Ahlers_2015}. The absence of a significant anisotropy is
consistent with the majority ($\gtrsim 85$ per cent) of the neutrino flux
being due to extragalactic population(s)
\citep[e.g.][]{Aartsen2017,Aartsen2019}.
So far, however, only one astronomical source has been significantly
associated (at the $3 - 3.5\,\sigma$ level) in time and space with IceCube
neutrinos, i.e. the blazar TXS\,0506+056 at $z=0.3365$
\citep{icfermi,iconly,padovani2018,Paiano_2018}. 
\cite{Aartsen2020a} have also reported an excess of neutrinos at the $2.9\,\sigma$ 
level from the direction of the local ($z = 0.004$) Seyfert 2 galaxy NGC\,1068 
and a $3.3\,\sigma$ excess in the northern sky due to significant p-values 
in the directions of NGC 1068 and three blazars: TXS\,0506+056, PKS\,1424+240, and 
GB6\,J1542+6129. 
Moreover, \cite{Stein_2021} have presented the association of a radio-emitting tidal 
disruption event (AT2019dsg) with a high-energy neutrino, giving a probability of finding 
such an event by chance $\sim 0.5$ per cent, which decreases to $\sim 0.2$ per cent 
($\sim 2.9\,\sigma$) if one takes into account its bolometric flux.
Very recently, \cite{Abbasi_2021} have performed a search for flare 
neutrino emission in the 10 years of IceCube data and reported a cumulative time-dependent 
neutrino excess in the northern hemisphere at the level of 3\,$\sigma$ associated with 
four sources: a radio galaxy, M87, two blazars (TXS\,0506+056 and 
GB6\,J1542+6129) and the Seyfert 2 galaxy (NGC 1068), these last three sources 
also being associated with neutrinos by \cite{Aartsen2020a}.
Blazars are a rare class
of Active Galactic Nuclei \citep[AGN; see][for
  reviews]{falomo2014,Padovani_2017} having a relativistic jet that is
pointing very closely to our line of sight
\citep[e.g.][]{UP95,Padovani_2017}. The case for a sub-class of blazars
being neutrino sources, however, is mounting \citep[see also][]{righi2019}.
\cite{giommi2020a} (hereafter G20), have carried out a detailed dissection of all
the public IceCube high-energy neutrinos that are well reconstructed 
(so-called tracks) and off the Galactic
plane, which provided a $3.2\,\sigma$ (post-trial) correlation
excess with $\gamma$-ray detected IBLs\footnote{Blazars are sub-divided
from a spectral energy distribution (SED) point of view on the basis of the
rest-frame frequency of their low-energy (synchrotron) hump (\nup) into
low- (LBL/LSP: \nup~$<10^{14}$~Hz [$<$ 0.41 eV]), intermediate- (IBL/ISP:
$10^{14}$~Hz$ ~<$ \nup~$< 10^{15}$~Hz [0.41 -- 4.1 eV)], and high-energy
(HBL/HSP: \nup~$> 10^{15}$~Hz [$>$ 4.1 eV]) peaked sources respectively
\citep{padgio95,Abdo_2010}. Extreme blazars are defined by \nup~$> 2.4
\times 10^{17}$~Hz [$>$ 1 keV] \citep[see, e.g.][for a recent review]{biteau2020}.} 
and HBLs. No excess was
found for LBLs. Given that TXS\,0506+056 is also a blazar of the IBL/HBL
type \citep{Padovani_2019} this result, together with previous findings,
consistently points to growing evidence for a connection between some
IceCube neutrinos and IBL and HBL blazars. We note that some papers 
have also found correlations with LBLs rather than IBLs/HBLs both through
statistical tests and studies of individual sources 
\citep[e.g.][we discuss these
and other papers in Section \ref{sec:detect}]{Kadler_2016,Hovatta_2021,Plavin_2020,Plavin_2021}.

Based on the results of G20, out of the 47 IBLs and HBLs in their Table 5, 
{$ 16\pm4$ could be new neutrino sources waiting to be identified. 
Further progress requires optical spectra, which are needed to measure
the redshift, and hence the luminosity of the source, determine the
properties of the spectral lines, and possibly estimate the mass of the
central black hole, $M_{\rm BH}$.

\cite{Paiano_2021} (hereafter Paper I) presented the spectroscopy of a
large fraction of the objects selected by G20, which, 
together with results taken from the literature, covered $\sim 80$ per cent
of that sample. Paper I is the first publication of the project entitled 
``The spectra of
IceCube Neutrino (SIN) candidate sources'' whose aim is threefold:
(1) determine the nature of the sources; (2) model their SEDs using all
available multi-wavelength data and subsequently the expected neutrino
emission from each blazar; (3) determine the likelihood of a connection 
between the neutrino and the blazar using a physical model for the 
blazar multi-messenger emissions, as done, for example, by
some of us in \cite{Petro_2015,Petro_2020}.

The purpose of this paper is to characterise the sources studied in Paper I, 
to determine their real nature, extending the work done by
\cite{Padovani_2019} for TXS\,0506+056 to a much larger sample of possible 
neutrino sources, and also see if these sources are any different from 
the rest of the blazar population. Given that only less than half of 
our sample is expected to be associated with IceCube tracks, this last point
might not be very straightforward. Although the paper is data-driven, we nevertheless
perform also a preliminary theoretical analysis of our sample.

The paper
is structured as follows: Section \ref{sec:sample} describes the sample we used, while
Section \ref{sec:ancillary} deals with ancillary data. In Section \ref{sec:characterisation} 
we present our main results on the source characterisation, Section \ref{sec:extra} 
deals with some additional sources, while Section \ref{sec:theory} gives a theoretical interpretation 
of our results in the context of high-energy neutrino emission.
Finally, Section \ref{sec:discussion} discusses our
results and Section \ref{sec:conclusions} summarises our conclusions. In Appendix 
\ref{sec:Appendix} we describe how we estimate $M_{\rm BH}$. 
We use a $\Lambda$CDM
cosmology with Hubble constant $H_0 = 70$ km s$^{-1}$ Mpc$^{-1}$, matter
density $\Omega_{\rm m,0} = 0.3$, and dark energy density
$\Omega_{\Lambda,0} = 0.7$. Spectral indices are defined by $S_{\nu}
\propto \nu^{-\alpha}$ where $S_{\nu}$ is the flux at frequency $\nu$.

\section{The sample}\label{sec:sample}

Paper I combined a literature search with newly obtained spectra and presented redshifts
for 36 IBLs and HBLs on the G20's list (see Tables 1 and 3 therein). These form the basis
of our sample, which also includes two sources, 5BZB J1322+3216 and 3HSP J152835.7+20042, 
for which we derived an estimate of the redshift as discussed in 
Appendix \ref{sec:Appendix} and from the host galaxy contribution to the SED \citep[see, e.g.][]{3HSP}, 
respectively. In the following, we refer 
to these two redshift estimates as ``photometric'' for brevity. To these, we further add
M87\footnote{The radio galaxy M87, one of the brightest objects
in the extragalactic sky, is well within the uncertainty region of the
neutrino track IceCube-141126A (G20: see also Section \ref{sec:Introduction}). As pointed out by G20 its
jet inclination and superluminal motion make it {\it almost} a blazar.
Moreover, M87 is HBL-like due its strong GeV and TeV emission combined with
a flat $\gamma$-ray spectrum and large flux variability, although the
complexity of its radio, optical, and near-infrared emission prevent us
from reliably estimating \nup.} and 3HSP J095507.9+35510
\citep{giommi2020b}, an extreme blazar associated with an IceCube track, 
which was announced after the 
G20 paper was completed but which still fulfils all the criteria for the sample construction 
of that paper. These 40 objects are all $\gamma$-ray detected and have been chosen by
selection to have $|b_{\rm II}|>10^{\circ}$ and (M87 excluded) \nup~$>10^{14}$~Hz. The
G20's sources are matched to 30 neutrino events out of 70 having an area
of the error ellipse smaller than that of a circle with radius r~$=
3^{\circ}$. More than one blazar is within the IceCube error ellipse in
most cases. 
Finally, Paper I presented also the spectra of some additional 
targets included in a preliminary version of the G20's list. These were still $\gamma$-ray 
emitting blazars without a redshift determination, which turned out not to fulfil all the 
final criteria adopted by those authors. In the following we refer to these sources as 
the ``extra'' sources and discuss those with \nup~$>10^{14}$~Hz} in Section \ref{sec:extra}. 
Given their ``additional'' nature these sources are not included in the 
statistical analysis of Section~\ref{sec:characterisation}.

\section{Ancillary data}\label{sec:ancillary}

\subsection{Radio data and \nup}

Radio flux densities at 1.4 GHz are given in Table 5 of G20
and span the range $0.003 - 1$ Jy. These come from radio catalogues and therefore 
represent typical values. \nup~values have been estimated on the basis of the 
multifrequency data from the catalogues and spectral databases called by the 
VOU\_Blazars tool V1.92 \citep{vou-blazar}, time averaged when multiple observations 
at the same frequency were present (see also Section 3.2 of G20). Since IBLs and HBLs 
are known to be highly variable both in flux and \nup, the estimation of the latter 
parameter, especially for those objects for which only sparse data are available, 
may change, even significantly, when additional data become available.  
Radio powers at 1.4 GHz, assuming a flat radio spectrum 
($\alpha_{\rm r} = 0$), and \nup~values (rest-frame, i.e. multiplied by $(1+z)$)
are given in Table \ref{tab:sample}. 

\subsection{Optical data}

Paper I has presented optical spectroscopy secured at the Gran
Telescopio Canarias and at the ESO Very Large Telescope for 17 sources in the
G20's sample with unknown redshift. The latter was determined for
nine objects, spanning the $0.09 - 1.6$ range, while for the others lower
limits were derived based either on the
absence of spectral signatures of the host galaxy or intervening absorption systems. Forbidden and
semi-forbidden emission lines with powers in the range $\sim 10^{40} - 4
\times 10^{41}$ erg s$^{-1}$ were also detected.

\subsubsection{Additional optical data}\label{sec:additional_optical}

Spectra for twelve additional sources were retrieved from the Sloan Digital
Sky Survey \citep[SDSS:][]{Ahumada2020} and ZBLLAC
database\footnote{\url{http://www.oapd.inaf.it/zbllac/}} and were used to
measure the line luminosity of, or an upper limit to, [\ion{O}{III}] 5007
\AA~and [\ion{O}{II}] 3727 \AA. The latter was derived by simulating the
emission features at the expected line position for a range of equivalent
widths, setting the upper limit based on the signal to noise of the
spectrum at the relevant wavelength. [\ion{O}{III}] and [\ion{O}{II}]
lines were detected in three and two objects respectively. All [\ion{O}{III}] 
and [\ion{O}{II}] line powers are given in Table \ref{tab:sample}.

\subsubsection{Disc luminosities for additional sources}\label{sec:Paliya}

For five sources for which we had no line information we got an upper limit
on $L_{\rm disc}$ from \cite{Paliya_2021}. This was used to derive an upper
limit on the accretion-related bolometric luminosity (Section \ref
{sec:characterisation}). These sources have no $L_{\rm [\ion{O}{II}]}$ and
$L_{\rm [\ion{O}{III}]}$ entries and an upper
limit in Table \ref{tab:sample} on the Eddington ratio, i.e, the ratio between the
(accretion-related) observed luminosity and the Eddington luminosity,
$L_{\rm Edd} = 1.26 \times 10^{46}~(M/10^8 \rm M_{\odot})$ erg s$^{-1}$,
where $\rm M_{\odot}$ is one solar mass.

\subsubsection{Black hole masses}\label{sec:masses}

The optical spectrum of our sources is typically made up of a non-thermal 
component superposed on the host galaxy spectrum \citep[see,
  e.g.][]{falomo2014}. When possible, we have therefore decomposed 
the observed optical
spectra into a power-law plus a template spectrum for the host galaxy
following \cite{mannucci2001}. The decomposition was obtained from a best
fit assuming two free parameters: the nucleus-to-host ratio and the
spectral slope of the non-thermal component. This
decomposition yields a good representation of the observed spectra, with a
ratio of the two components, evaluated at 6500 \AA, ranging from 0.3 to 5
(with a median value of 2.5). From the observed magnitude of the host galaxy in the
R band we then computed the absolute magnitude $M(R)$ including
the k-correction and a correction for passive (i.e. with the bulk of the stars 
formed at high redshift) evolution \citep{bressan1998} 
in order to properly match the observations with local Universe data. 
Finally, we estimated $M_{\rm BH}$ for these sources using the 
$M_{\rm BH} - M(R)$ relationship of \cite{labita07} (consistent with 
the one by \cite{McLure_2002} for the same cosmological parameters), 
which gives a typical uncertainty $\approx 0.45$ dex (we refer the reader 
to Appendix \ref{sec:Appendix} for more details). The mass for M87 is from 
\cite{EHT_2019}, the upper limit on the mass of TXS 0506+065 assumes 
the host galaxy to be a typical giant elliptical with $M(R) \sim -22.9$, 
or fainter \citep[see discussion in][]{Padovani_2019}, while for CRATES J232625+011147 
we use the mass obtained by \cite{Paliya_2021} by applying the virial technique to the 
\ion{Mg}{II} line. For the 35 per cent of the sources for which we could not 
estimate $M_{\rm BH}$ we adopt $M_{\rm BH} = 6.3 \times 10^8 M_{\odot}$,
which assumes the host galaxy to be a typical giant elliptical \citep{labita07}, 
a value which is very close to the median of our masses ($M\sim 4 
\times 10^8 M_{\odot}$). $M_{\rm BH}$ values are reported in Table \ref{tab:sample} 
and detailed in Table \ref{tab:apdx}.

\subsection{$\gamma$-ray data}

The analysis of the sources' $\gamma$-ray emission is based on {\tt Fermipy} 
\citep{2017ICRC...35..824W} and uses publicly available {\it Fermi}-LAT Pass 8 data (with 
the \texttt{P8R3\_SOURCE\_V3} instrument response function) acquired in the 
period August 4, 2008 to mid 2020 and follows the standard procedures suggested 
by the {\it Fermi}-LAT team. Namely, we select only events with a high 
probability of being photons (evclass = 128, evtype = 3 [FRONT+BACK]) in 
the energy range between 100\,MeV and 300\,GeV from a region of interest (ROI) of 
20$^\circ$ around the source's position. Possible contamination from the Earth limb is removed by cutting 
out events with zenith angle $> 90^\circ$ and only considering time intervals in 
which the data acquisition of the spacecraft was stable (DATA\_QUAL>0 \&\& LAT\_CONFIG==1). 

Each fit is based on an emission model that, besides the target source at 
its multi-frequency position, incorporates the isotropic (\texttt{gll\_iem\_v07}) 
and Galactic (\texttt{iso\_P8R3\_SOURCE\_V3\_v1}) diffuse component, as well as all 
the known $\gamma$-ray point sources in the region. In addition to the spectral parameters 
of the target source the flux normalisation of the diffuse component and the spectral parameters 
of all point and extended sources within $\sim$ 13$^{\circ}$ (95 per cent point spread function 
at 100 MeV) are treated as free parameters. This procedure thereby efficiently reduces the risk of source 
confusion. Finally, the best-fit flux of the target source is determined by maximising the signal 
likelihood $\mathcal{L}(\textrm{full\,model | data})$. For the significance we further maximise the 
background likelihood $\mathcal{L}(\textrm{bkg\,model | data})$, i.e. the model without the 
target source. The test-statistic is then defined as
\begin{equation}
    TS = 2\times \left[ \log\mathcal{L}(\textrm{full\,model | data}) - \log \mathcal{L}(\textrm{bkg\,model | data})\right].
\end{equation}
The TS value can be converted to a significance using the $\chi^2$ distribution with 
two-degrees of freedom. We can then confirm the significant 
detection (well $>3\,\sigma$) of all but two sources in our sample independently of the 
{\it Fermi}-LAT Fourth Source Catalog (4FGL: \citealt{4FGL}). The remaining two objects, 
i.e. 3HSP J062753.3$-$15195 (3FGL J0627.9$-$1517) and 3HSP J125821.5+21235 (3FGL J1258.4+2123), 
have dropped to 2.1 and 2.5$\,\sigma$ for the time-integrated emission (down from $\sim 4.5$ 
and 5.0$\,\sigma$, respectively) and are not included in the {\it Fermi-}4FGL-DR2 catalogue \citep{4FGL-DR2}. 
This drop in significance is characteristic for strongly time-variable objects and therefore 
not unexpected. Furthermore, visual inspection shows that the multi-wavelength SEDs are 
fully-consistent with the derived $\gamma$-ray spectra. Rest-frame, k-corrected, 
$\gamma$-ray powers between 0.1 and 100 GeV, $L_{\gamma}$, were then derived by 
integrating the best-fit spectra, together with  $1\,\sigma$ statistical uncertainties. 

\begin{table*}
\caption{Main sample properties.}
 \begin{center}
 \begin{tabular}{lllrrrrrrr}
   \hline
    Name & 4FGL name & \magg$z$~~& \nup & $P_{\rm 1.4GHz}$ & $L_{\rm [\ion{O}{II}]}$ & $L_{\rm [\ion{O}{III}]}$ & $M_{\rm BH}$ & $L/L_{\rm Edd}$ & $L_{\gamma}/L_{\rm Edd}$ \\
         &  &  & [Hz] & [W Hz$^{-1}$] & [erg s$^{-1}$] & [erg s$^{-1}$] & [$M_{\odot}$] &  &  \\
      \hline
 {\bf 3HSP J010326.0+15262} & 4FGL~J0103.5+1526 & \magg0.2461 &  15.1 &  25.52 & <40.8 &  40.9 &  8.6 &  -1.8 &  -2.1 \\
 {\bf 5BZU J0158+0101}     & 4FGL~J0158.8+0101 & \magg0.4537 &  14.3 &  25.63 & <41.0 &  40.7 &  7.4 &  -0.8 &  -0.1 \\
 VOU J022411+161500   & 4FGL~J0224.2+1616 & >0.5 &  14.7 & >24.93 &   ...~~~  &   ...~~~  &  ...~~~&   ...~~~  & >-1.5 \\
 {\it 3HSP J023248.5+20171} & 4FGL~J0232.8+2018 & \magg0.139 &  18.6 &  24.58 &   ...~~~  &   ...~~~  &  8.8 & <-3.4 &  -2.7 \\
 3HSP J023927.2+13273 & 4FGL~J0239.5+1326 & >0.7 &  15.2 & >25.41 &   ...~~~  &   ...~~~  &  ...~~~&   ...~~~  & >-1.4 \\
 CRATES J024445+132002 & 4FGL~J0244.7+1316 & \magg0.9846 &  14.8 &  26.55 &  41.6 &   ...~~~&  ...~~~&  -1.3 &  -0.6 \\
{\it 3HSP J033913.7$-$17360} & 4FGL~J0339.2$-$1736 & \magg0.066 &  15.6 &  24.23 &   ...~~~  &   ...~~~  &  8.6 & <-3.0 &  -2.8 \\
 3HSP J034424.9+34301 & 4FGL~J0344.4+3432 & >0.25 &  15.8 & >24.29 &   ...~~~  &   ...~~~  &  ...~~~&   ...~~~  & >-2.4 \\
 {\bf TXS 0506+056} & 4FGL~J0509.4+0542 & \magg0.3365 &  14.6 &  26.18 &  41.3 &  41.3 &  <8.8 &  >-1.7 &  >-0.4 \\
 {\bf CRATES J052526$-$201054} & 4FGL~J0525.6$-$2008 & \magg0.0913 &  14.5 &  24.64 &   ...~~~&  40.3 &  8.1 &  -1.7 &  -2.7 \\
{\it 3HSP J062753.3$-$15195} & 3FGL~J0627.9$-$1517 &  \magg0.3102 &  17.4 &  25.01 & <41.1 & <40.8 &  9.0 & <-2.0 &  -2.6 \\
 3HSP J064933.6$-$31392 & 4FGL~J0649.5$-$3139 & >0.7 &  17.2 & >24.98 &   ...~~~  &   ...~~~  &  0.0 &   ...~~~  & >-1.0 \\
{\it 3HSP J085410.1+27542} & 4FGL~J0854.0+2753 &  \magg0.493 &  16.3 &  24.96 &   ...~~~  &   ...~~~  &  9.0 & <-2.5 &  -2.5 \\
{\it 3HSP J094620.2+01045} & 4FGL~J0946.2+0104 &  \magg0.576 & >18.2 &  25.11 & <41.7 & <40.5 &  8.6 & <-1.6 &  -1.0 \\
  3HSP J095507.9+35510 & 4FGL~J0955.1+3551 &  \magg0.557 &  17.9 &  24.81 & <40.3 & <40.3 &  8.8 & <-2.0 &  -1.7 \\
 {\bf GB6 J1040+0617}       & 4FGL~J1040.5+0617 &  \magg0.74 &  14.7 &  25.70 &  41.0 &   ...~~~&  ...~~~&  -1.5 &  -0.2 \\
 5BZB J1043+0653      & 4FGL~J1043.6+0654 & >0.7 &  14.7 & >25.00 &   ...~~~  &   ...~~~  &  ...~~~&   ...~~~  & >-1.3 \\
 {\bf 3HSP J111706.2+20140} & 4FGL~J1117.0+2013 &  \magg0.138 &  16.6 &  24.66 &  40.7 & <40.4 &  8.1 &  -0.9 &  -1.3 \\
 M87                  & 4FGL~J1230.8+1223 &  \magg0.0043 &   ...~~~&  $^a$22.60 &   ...~~~&  ...~~~&  $^b$9.8 & <-6.2 &  -6.1 \\
 3HSP J123123.1+14212 & 4FGL~J1231.5+1421 &  \magg0.2558 &  16.1 &  24.95 & <41.2 & <40.9 &  8.4 & <-1.4 &  -1.7 \\
 3HSP J125821.5+21235 & 3FGL~J1258.4+2123 &  \magg0.6265 &  16.9 &  25.43 & <41.1 &   ...~~~&  8.0 & <-0.7 &  -1.2 \\
 3HSP J125848.0$-$04474 & 4FGL~J1258.7$-$0452 &  \magg0.4179 &  17.2 &  24.28 & <40.5 & <40.3 &  8.4 & <-1.6 &  -1.3 \\
 3HSP J130008.5+17553 & 4FGL~J1300.0+1753 & >0.6 &  14.7 & >25.18 &   ...~~~  &   ...~~~  &  ...~~~&   ...~~~  & >-1.6 \\
 5BZB J1314+2348      & 4FGL~J1314.7+2348 & \magg0.15$^c$ & >14.1 & 24.99 &   ...~~~  &   ...~~~  &  ...~~~&   ...~~~  & -2.1 \\
 {\it 5BZB J1322+3216}      & 4FGL~J1321.9+3219 &  \magg0.4$^d$ &  14.6 &  26.56 &   ...~~~  &   ...~~~  &  8.2 & <-2.1 &  -1.4 \\
{\it VOU J135921$-$115043}   & 4FGL~J1359.1$-$1152 &  \magg0.242 &  14.1 &  24.83 &   ...~~~& <40.3 &  8.6 & <-2.1 &  -2.7 \\
 3HSP J140449.6+65543 & 4FGL~J1404.8+6554 &  \magg0.3627 &  16.1 &  24.71 & <41.0 & <40.8 &  8.3 & <-1.3 &  -1.1 \\
{\it  VOU J143934$-$252458}   & 4FGL~J1439.5$-$2525 &  \magg0.16 &  14.1 &  24.32 &   ...~~~  &   ...~~~  &  8.6 & <-3.1 &  -2.6 \\
 {\bf 3HSP J143959.4$-$23414} & 4FGL~J1440.0$-$2343 &  \magg0.309 &  16.3 &  25.38 &  40.9 &  40.5 &  8.6 &  -1.7 &  -1.6 \\
 3HSP J144656.8$-$26565 & 4FGL~J1447.0$-$2657 &  \magg0.3315 &  17.7 &  25.05 & <40.5 & <40.4 &  8.6 & <-1.8 &  -2.0 \\
 3HSP J152835.7+20042 & 4FGL~J1528.4+2004 &  \magg0.52$^d$ &  16.4 &  24.39 &   ...~~~  &   ...~~~  &  ...~~~&   ...~~~  &  -1.8 \\
 3HSP J153311.2+18542 & 4FGL~J1533.2+1855 &  \magg0.307 &  17.1 &  24.73 & <41.3 & <40.9 &  8.4 & <-1.4 &  -1.6 \\
 3HSP J155424.1+20112 & 4FGL~J1554.2+2008 &  \magg0.2223 &  17.4 &  24.98 & <40.9 & <40.5 &  8.6 & <-1.8 &  -2.2 \\
 3HSP J180849.7+35204 & 4FGL~J1808.8+3522 &  \magg0.141 &  15.1 &  24.15 & <40.5 & <40.3 &  8.0 & <-1.2 &  -2.3 \\
 3HSP J213314.3+25285 & 4FGL~J2133.1+2529 &  \magg0.294 &  15.3 &  24.93 &   ...~~~  &   ...~~~  &  8.8 &   ...~~~  &  -2.0 \\
 3HSP J222329.5+01022 & 4FGL~J2223.3+0102 & >0.7 &  15.7 & >24.94 &   ...~~~  &   ...~~~  &  ...~~~&   ...~~~  & >-1.4 \\
 {\bf 5BZB J2227+0037}     & 4FGL~J2227.9+0036 & >1.0935 &  14.8 & >26.50 &   ...~~~  &   ...~~~  &  ...~~~&   ...~~~  & >-0.2 \\
 {\bf CRATES J232625+011147} & 4FGL~J2326.2+0113 &  \magg1.595 &  14.4 &  27.12 &   ...~~~ &   ...~~~ &  9.3 &  -2.7 &  -0.7 \\
 3HSP J235034.3$-$30060 & 4FGL~J2350.6$-$3005 &  \magg0.2328 &  15.8 &  24.70 &   ...~~~  &   ...~~~  &  ...~~~&   ...~~~  &  -2.1 \\
 IC 5362              & 4FGL~J2351.4$-$2818 &  \magg0.0276 &  14.5 &  23.19 &   ...~~~  &   ...~~~  &  ...~~~&   ...~~~  &  -4.6 \\
  \hline
   \end{tabular}
   \label{tab:tab1}
  \end{center}
\footnotesize \textit{Notes.} All values, apart from redshift, are in logarithmic scale; $^a$time-averaged VLBA core power at 15 GHz \protect\citep{Kim_2018};\\
$^b$\cite{EHT_2019}; $^c$uncertain redshift; $^d$photometric redshift. Masquerading BL Lacs are 
indicated in {\bf bold face}, while non-masquerading ones in {\it italics}.
 
\label{tab:sample}
\end{table*}

\section{Source characterisation}\label{sec:characterisation}

We now characterise our sources using the data discussed above. The main
issue we want to address is that of the so-called ``masquerading'' BL
Lacs. \cite{Padovani_2019} showed that TXS\,0506+056, the first plausible
non-stellar neutrino source is, despite appearances, {\it not} a blazar of
the BL Lac type but instead a masquerading BL Lac. This class was
introduced by \cite{giommibsv1,giommibsv2} (see also
\citealt{Ghisellini_2011}) and includes sources which are in reality 
FSRQs\footnote{Blazars have always been divided from an optical spectroscopy 
point of view in two classes, i.e. flat-spectrum radio quasars (FSRQs) 
and BL Lac objects. FSRQs display broad, quasar-like strong emission lines 
while BL Lacs often are totally featureless and sometimes exhibit weak 
absorption and emission lines \citep{UP95}. Most FSRQs are LBLs, with only a  
very small fraction of them being IBLs.}
whose emission lines are washed out by a very bright, Doppler-boosted jet
continuum, unlike ``real'' BL Lacs, which are {\it intrinsically}
weak-lined objects. This is relevant because ``real'' BL Lacs and FSRQs
belong to different physical classes, i.e. objects {\it without} and {\it
  with} high-excitation emission lines in their optical spectra, referred
to as low-excitation (LEGs) and high-excitation galaxies (HEGs),
respectively\footnote{The LEG/HEG classification applies to all AGN, with
quasars and Seyferts belonging to the HEG category and low-ionisation
nuclear emission-line regions (LINERs) and absorption line galaxies being
classified as LEGs: see, e.g.  \cite{Padovani_2017}.}. TXS 0506+056, being
a HEG, therefore benefits from several radiation fields external to the jet
(i.e. the accretion disc, photons reprocessed in the broad-line region or
from the dusty torus), which, by providing more targets for the protons
might enhance neutrino production as compared to LEGs. This makes
masquerading BL Lacs (and, at first glance, FSRQs: see Section 
\ref{sec:detect}) particularly attractive from the point of view of
high-energy neutrinos.

As per \cite{Padovani_2019}, to which we refer the reader for more details, 
we use the following parameters to look for masquerading BL Lacs, i.e. BL Lacs 
with HEG-like (i.e. quasar-like) properties, in our sample, in decreasing order 
of relevance: 

\begin{enumerate}
    \item the location on the radio power -- emission line power, $P_{\rm 1.4GHz}$ -- 
      $L_{\rm [\ion{O}{II}]}$, 
      diagram \citep[Fig. 4 of][]{Kalfountzou_2012}, which defines the
      locus of jetted (radio-loud) quasars. Any BL Lac falling on this locus 
      has to be a masquerading one, while real BL Lacs will occupy the lower-left 
      part of the plot. This is quite straightforward 
      but requires the detection (or an upper limit on the flux) of the
      [\ion{O}{II}] 3727 \AA~line; 
    \item a radio power $P_{\rm 1.4GHz} > 10^{26}$ W Hz$^{-1}$, since HEGs become the
      dominant population in the radio sky above this value
      \citep{hec14}. Although this is strictly valid for the local
      Universe, it can be applied to higher redshifts as well because 
      LEGs do not evolve much in luminosity, unlike HEGs
      \citep[e.g.][]{Padovani_2015a}. 
      Note that there are many HEGs below this limit so the fact
      that a source does not satisfy this condition does not rule out 
      a masquerading BL Lac classification;
    \item an Eddington ratio $L/L_{\rm Edd}
      \gtrsim 0.01$, which is typical of HEGs. The derivation of the
      thermal, accretion-related bolometric luminosity is not that easy 
      for BL Lacs and it involves measuring $L_{\rm [\ion{O}{II}]}$
      and $L_{\rm [\ion{O}{III}]}$ \citep[e.g.][]{Punsly_2011}. In the case
      of CRATES J232625+011147, at $z=1.595$, we used its \ion{Mg}{II} and
      \ion{C}{III]} luminosities to estimate the broad-line region (BLR)
      luminosity, $L_{\rm BLR}$, from the composite spectrum of
      \cite{van01} (we then derived $L_{\rm disc} \sim 10 \times L_{\rm
        BLR}$ assuming a standard covering factor of $\sim 10$ per cent and
      $L \sim 20 \times L_{\rm BLR}$: see \citealt{Padovani_2019}). The upper limit 
      for M87 has been derived from its limit on $L_{\rm BLR}$ \citep{Sbarrato_2014}, while the
      upper limits on $L_{\rm disc}$ from \cite{Paliya_2021} (Section \ref{sec:Paliya}) 
      turned into upper limits on $L < 2 \times L_{\rm disc}$ \citep{Padovani_2019}.
      $L_{\rm Edd}$ requires an estimate of the black hole mass, which is
      not straightforward to get (Section \ref{sec:masses});
   \item a $\gamma$-ray Eddington ratio $L_{\gamma}/L_{\rm Edd} \gtrsim 0.1$, 
     following \cite{Sbarrato_2012} who have
     proposed a division between ``real'' BL Lacs and FSRQs at this
     value. As is the case for $L/L_{\rm Edd}$ this also requires an estimate of the
     black hole mass. 
     
     \cite{Padovani_2019} derived also $L_{\rm
       BLR}/L_{\rm Edd}$, which was needed to estimate the $\gamma$-ray
     attenuation due to the BLR photon field. However, for BL Lacs the
     estimation of $L_{\rm BLR}$ depends on $L_{\rm [\ion{O}{II}]}$ and
     $L_{\rm [\ion{O}{III}]}$, as is the case for the accretion-related
     bolometric luminosity, and therefore this is not an independent
     parameter to characterise a source. Nonetheless, we do use $L_{\rm BLR}$ 
     in one of our theoretical scenarios in Section \ref{sec:ScenarioB}.
\end{enumerate}

We apply these four criteria in descending order of importance. We anticipate that 
no object is classified as masquerading based only on $L/L_{\rm Edd}$ and/or 
$L_{\gamma}/L_{\rm Edd}$, which are the least certain parameters given their $L$ and 
$M_{\rm BH}$ dependence. In practice, then, the Eddington ratios are used only as
a consistency check.  

\begin{table*}
 \caption{Masquerading BL Lac properties: main sample.}
 \begin{center}
 \begin{tabular}{lclcccc}
   \hline
    Name & IceCube Name & ~~~~~$z$ & $P_{\rm 1.4GHz} - L_{\rm \ion{O}{II}}$ & $P_{\rm 1.4GHz}$ & $L/L_{\rm Edd}$ & $L_{\gamma}/L_{\rm Edd}$ \\
      \hline
 3HSP J010326.0+15262  & IC160331A &  \magg0.2461 & \checkmark & I  & \checkmark & \xmark \\
 5BZU J0158+0101       & IC090813A &  \magg0.4537 & \checkmark & I  & \checkmark & \checkmark \\
 TXS 0506+056           & IC170922A &  \magg0.3365 & \checkmark & \checkmark & \checkmark & \checkmark \\
 CRATES J052526$-$201054 & IC150428A &  \magg0.0913 & \checkmark &  I & \checkmark & \xmark  \\
 GB6 J1040+0617         & IC141209A &  \magg0.74 & \checkmark & I  & \checkmark & \checkmark \\
3HSP J111706.2+20140  & IC130408A &  \magg0.1380 & \checkmark & I  & \checkmark & \xmark  \\ 
 3HSP J143959.4$-$23414 & IC170506A &  \magg0.309 & \checkmark &  I & \checkmark & \xmark  \\
 5BZB J2227+0037        & IC140114A & >1.0935 & --  & \checkmark & --  & \checkmark \\
 CRATES J232625+011147   & IC160510A &  \magg1.595 & -- & \checkmark & \xmark  & \checkmark \\
  \hline
  \end{tabular}
  \end{center}
\footnotesize {\textit{Notes.} ``I'' implies that the condition is not met but this does not mean this is not a masquerading BL Lac, --'' that no information is available.}
 \label{tab:masq}
\end{table*}


Table \ref{tab:sample} gives the main properties of the sample, namely name
(column 1), {\it Fermi-}4FGL  name (column 2), redshift (column 3), 
rest-frame \nup~(column 4), $P_{\rm 1.4GHz}$ (column
5), $L_{\rm [\ion{O}{II}]}$ and $L_{\rm [\ion{O}{III}]}$ (columns 6 and 7),
$M_{\rm BH}$ (column 8), $L/L_{\rm Edd}$ and $L_{\gamma}/L_{\rm Edd}$
(columns 9 and 10). When $M_{\rm BH}$ is missing we assumed a value of $6.3 \times 10^8
M_{\odot}$, as discussed above. Masquerading BL Lacs are indicated in bold face, 
non-masquerading ones in italics, while sources for which we do not have the 
relevant information to make a decision are in roman (Section \ref{sec:fract_masq}).
More details on masquerading BL Lacs are given in Table \ref{tab:masq}.\footnote{
Blazar variability is very unlikely to affect our source characterisation, for various 
reasons: (1) not a single source has been classified as masquerading BL Lac only on the 
basis of its $P_{\rm 1.4GHz}$ or its $L_{\gamma}/L_{\rm Edd}$; (2) IBLs and HBLs are known 
to be less $\gamma$-ray variable than LBLs \citep[e.g.][]{Rajput_2020}; (3) $\gamma$-ray 
variability is very luminosity dependent and becomes relevant at $L_{\gamma} 
\gtrsim 10^{46}$ erg $^{-1}$ \citep{Ackermann_2011b} where we have only a few sources 
(albeit all masquerading: but see points (1) and (2)).} 
We refer the reader to Tables 1 and 3 of
Paper I for redshift references and details on the spectra. Note that all
but one redshift lower limits have been derived from the lack of detection 
of host galaxy absorption lines assuming a standard elliptical host galaxy 
with $M(R) = -22.9$, the remaining limit (for 5BZB J2227+0037) being instead 
based on the detection of intervening absorption systems attributed to Mg~II 
and Fe~II.   

\subsection{The radio power -- emission line diagram ($P_{\rm 1.4GHz}$ -- 
$L_{\rm [\ion{O}{II}]}$)}\label{sec:P14_LOII}

\begin{figure}
\vspace{-2.2cm}
\hspace{-0.6cm}
\includegraphics[width=0.55\textwidth]{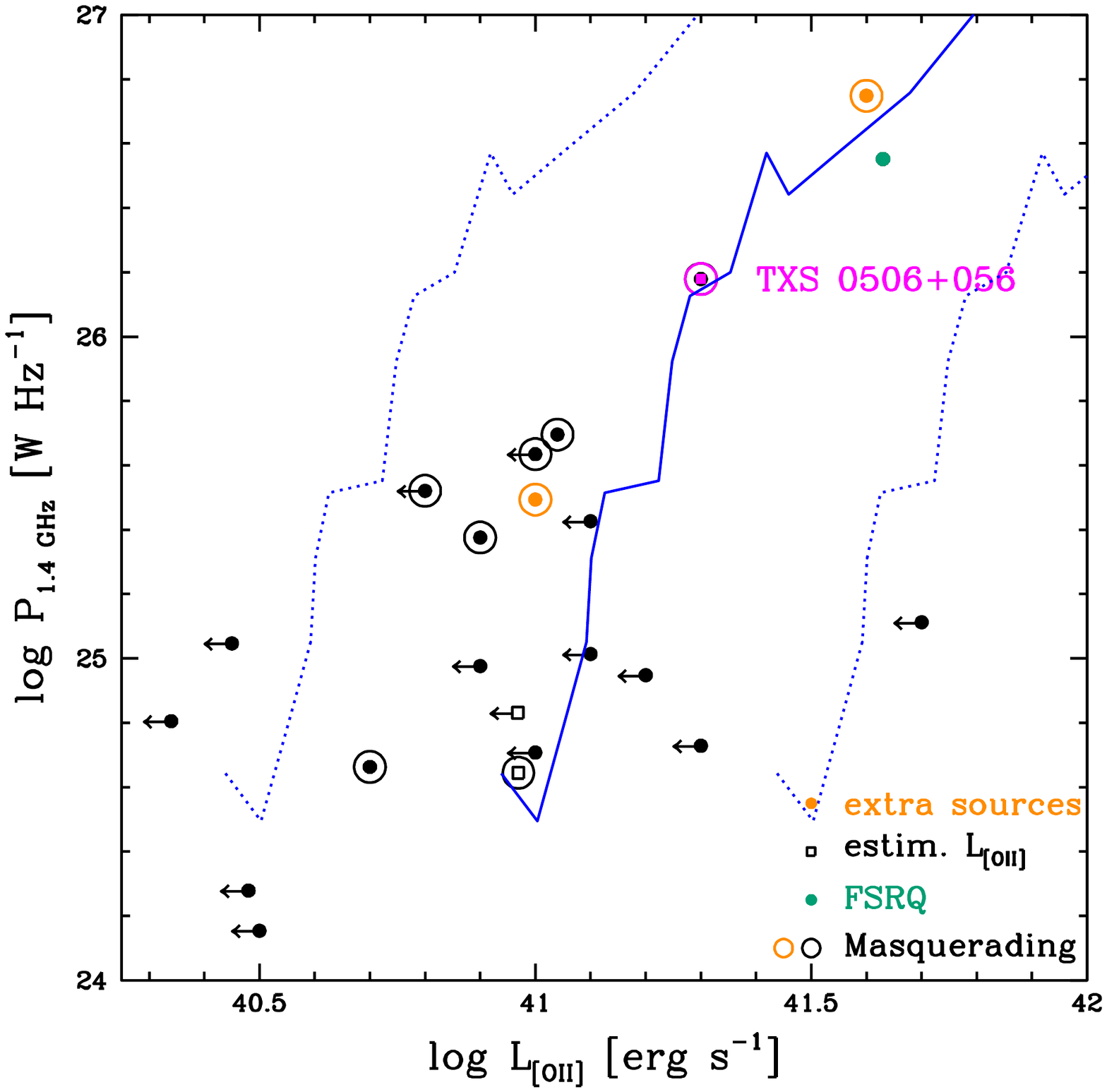}
\vspace{-3.1cm}
\caption{$P_{\rm 1.4GHz}$ vs. $L_{\rm [\ion{O}{II}]}$ for our sample (black
  filled circles), with masquerading sources highlighted (larger empty
  circles). Sources for which $L_{\rm [\ion{O}{II}]}$ has been estimated
  from $L_{\rm [\ion{O}{III}]}$ are denoted by black empty squares. The
  green filled circle indicates the single FSRQ in our sample, while orange points are 
  the extra sources. TXS\,0506+056, which also belongs to our sample, is indicated
  by a magenta circle. The solid blue line is the locus of jetted (radio-loud) quasars, with
  the two dotted lines indicating a spread of 0.5 dex, which includes most
  of the points in Fig. 4 of \protect\cite{Kalfountzou_2012} (converted from 
  radio powers in W Hz$^{-1}$ sr$^{-1}$ and line powers in W). 
  Arrows denote upper limits on $L_{\rm [\ion{O}{II}]}$.}
\label{fig:Lr_LOII}
\end{figure}

Fig. \ref{fig:Lr_LOII} shows the location of the sources with [\ion{O}{II}]
information on the $P_{\rm 1.4GHz}$ -- $L_{\rm [\ion{O}{II}]}$ diagram. To
increase our statistics we have also added two objects for which only the
[\ion{O}{III}] flux was available, converting $L_{\rm [\ion{O}{III}]}$ to
$L_{\rm [\ion{O}{II}]}$ using the scaling relation $L_{\rm [\ion{O}{II}]} 
\propto L_{\rm [\ion{O}{III}]}^{0.04}$, as implied by 
Fig. 7 of \cite{Kalfountzou_2012} (see the red
line therein; these sources are marked differently in the
figure). Seven objects are close to the locus of jetted AGN and are therefore
``bona fide'' masquerading BL Lacs. These include, apart from TXS
0506+056, 3HSP J010326.0+15262, 5BZU J0158+0101, CRATES J052526$-$201054,
GB6 J1040+0617, 3HSP J111706.2+20140, and 3HSP J143959.4$-$23414. All have
also $L/L_{\rm Edd} \ge 0.01$. Two of these objects (3HSP J010326.0+15262 and 5BZU
J0158+0101) have upper limits on their $L_{\rm [\ion{O}{II}]}$, although
quite close to the locus, but we still classify them as masquerading
because they have an [\ion{O}{III}] detection and $L/L_{\rm Edd} \sim
0.014$ and $\sim 0.17$, with the latter source having also
$L_{\gamma}/L_{\rm Edd} \sim 0.8$. CRATES J024445+132002, the only FSRQ in
our sample, is also (by definition) very close to the locus. Four more
sources have quite stringent $L_{\rm [\ion{O}{II}]}$ upper limits, while
eight more have an upper limit not too far from (or to the right of) the 
locus. However, given
that these latter sources have no other HEG-like property and no
[\ion{O}{III}] detection, and that we want to keep the selection conservative,
we are not including them with the masquerading sources.

\begin{figure}
\vspace{-2.2cm}
\hspace{-0.6cm}
\includegraphics[width=0.55\textwidth]{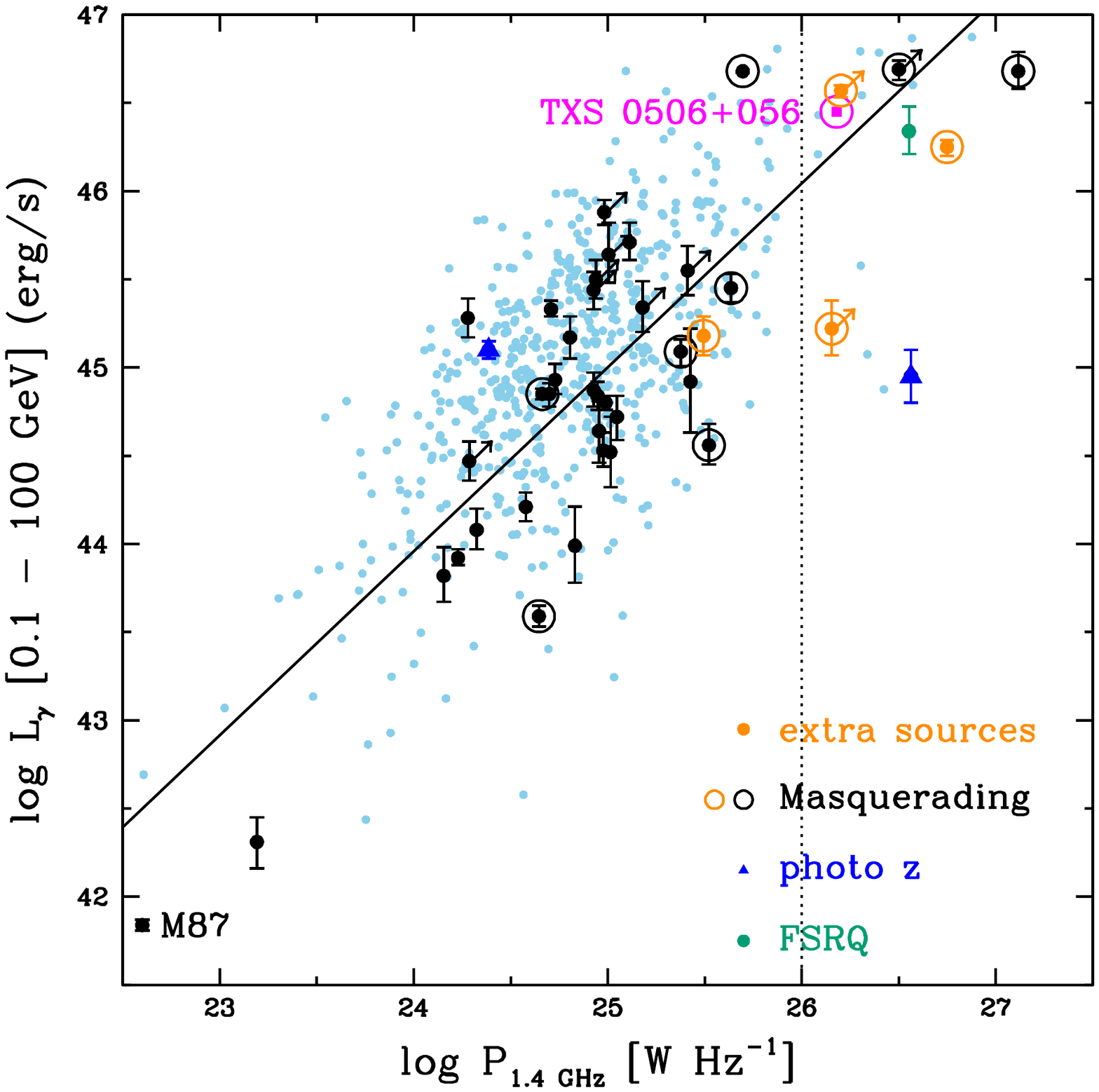}
\vspace{-3.1cm}
\caption{$L_{\gamma}$ vs. $P_{\rm 1.4GHz}$ for our sample (black filled
  circles), with masquerading sources highlighted (larger empty
  circles). The green filled circle indicates the single FSRQ, orange
  points are the extra sources, while the blue triangles denote sources for
  which we derived an estimate of the redshift from the host galaxy
  contribution to the SED \citep[see, e.g.][]{3HSP}. The radio power for
  M87, which is labelled, is derived from a time-averaged VLBA core flux at
  15 GHz \protect\citep{Kim_2018}. The average value for TXS\,0506+056
  (magenta point: \citealt{Padovani_2019}), which also belongs to our sample, 
  is also shown. Error bars denote the uncertainties. The solid line is the
  linear best fit $L_{\gamma} \propto P_{\rm 1.4GHz}^{1.04}$, while the 
  vertical dotted line marks the $10^{26}$ W Hz$^{-1}$ power above which
  a source is classified as masquerading (see text for details). 
  Arrows denote 
  lower limits on redshift and therefore powers. The
  small light blue points are the control sample of IBLs and HBLs (see
  Section \ref{sec:diff_blazar} for details).}
\label{fig:Lr_Lgamma}
\end{figure}

\subsection{The $\gamma$-ray power -- radio power diagram 
($L_{\gamma} - P_{\rm 1.4GHz}$)}\label{sec:P14_Lgamma}

Fig. \ref{fig:Lr_Lgamma} shows $L_{\gamma}$ vs. $P_{\rm 1.4GHz}$ for our
sources (black circles). We also include the time-averaged very long baseline array (VLBA)
core power at 15 GHz for M87 \citep{Kim_2018}, which, given its
substantially flat radio spectrum between $15 - 129$ GHz, should be
representative of its 1.4 GHz core power as  well. Two more sources get
classified as masquerading BL Lacs thanks to their $P_{\rm 1.4GHz} >
10^{26}$ W Hz$^{-1}$, namely 5BZB J2227+0037 and CRATES
J232625+011147. Both of them have also $L_{\gamma}/L_{\rm Edd} \gtrsim
0.6$. Note that 5BZB J1322+3216 (a.k.a. 4C+32.43, the rightmost blue
triangle) would also have fulfilled (only) the $P_{\rm 1.4GHz}$ criterion
but its radio spectrum is very steep ($\alpha_{\rm r} \sim 0.8$ between 130
MHz and 4.8 GHz) and not blazar-like. Its large radio power is therefore
due to extended emission and not to the core. This fits with the fact that
the host galaxy component is quite visible in its SED and suggests this is
a moderately beamed blazar with its non-thermal emission swamped by the
galaxy in the radio and near-IR bands \citep[see][]{giommibsv1}. All other
sources have flat radio spectra. Table \ref{tab:masq} lists our final list of 
masquerading BL Lacs with the conditions they fulfil; all sources satisfy at least
two criteria.

Fig. \ref{fig:Lr_Lgamma} shows a very strong linear correlation between the
two powers, significant at the $> 99.99$ per cent level\footnote{We exclude
from the fit the FSRQ, M87, and 5BZBJ 1322+3216 for reasons discussed
above but this has no influence on the significance or the slope.} and 
with $L_{\gamma} \propto
P_{\rm 1.4GHz}^{1.04\pm0.13}$. (We can exclude that this strong correlation 
is due to the common redshift dependence of the two powers, as its removal 
by using a partial correlation
analysis \citep[e.g.][]{Padovani_1992} still gives a 99.8 per cent level
significance.) Note that the sample includes also eight
sources with lower limits on their redshifts and therefore on their
powers. Since the application of survival analysis in this case requires
binning, the resulting fit depends on the chosen bin size. We have then
tested how this affects our results by increasing artificially the lower
limits on the powers by 0.75 dex, which corresponds, for example, to a
(large) increase from $z=0.7$ to $z=1.4$. The new best fit is $L_{\gamma}
\propto P_{\rm 1.4GHz}^{1.14\pm0.11}$ with the same significance, fully
consistent with our previous results. The issue of how the properties
of masquerading sources might differ from those of the other sources is 
discussed in Section \ref{sec:masq_diff}. 

This correlation is not surprising as previous studies have noted the
strong $\gamma$-ray -- radio correlation for HBLs, stronger than for other
blazar sub-classes \citep[e.g.][]{Ackermann_2011a,Lico_2017}, and $\sim 62$
per cent of our sources are HBLs. Furthermore, one should expect a  
correlation between radio and $\gamma$-ray emission in IBLs/HBLs given that
their SED is generally consistent with simple one-zone synchrotron self-Compton 
models with roughly constant ratio between the peak of the 
Compton and \nup~luminosity, the so-called Compton dominance 
\citep{Abdo_2010,GiommiPlanck}. Given that \nup~in 
these sources is located close to the optical band or at higher energies, the ratio 
between optical and $\gamma$-ray fluxes is close to the Compton dominance, which is 
similar in most objects. Since the radio to optical spectral slope ($\alpha_{\rm ro}$) 
in these sources is also known to be 
approximately constant \citep{padgio95,Abdo_2010} it then follows that the radio and 
$\gamma$-ray powers are correlated.

\subsection{The synchrotron peak -- $\gamma$-ray power diagram 
(\nup~-- $L_{\gamma}$)}\label{sec:sequence}

\begin{figure}
\vspace{-2.2cm}
\hspace{-0.6cm}
\includegraphics[width=0.55\textwidth]{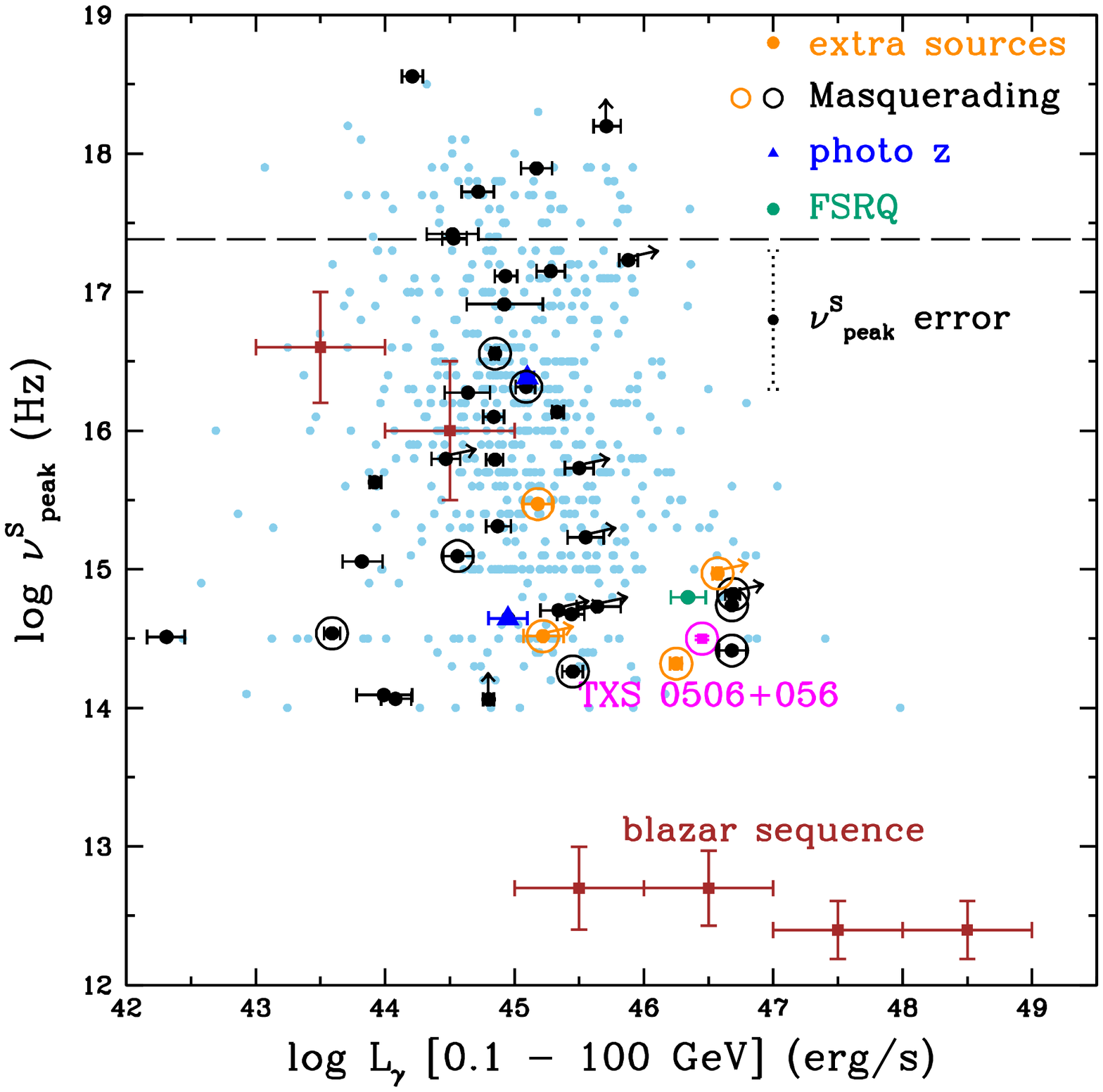}
\vspace{-3.1cm}
\caption{\nup~vs. $L_{\gamma}$ for our sample (black filled circles),
  with masquerading sources highlighted (larger empty circles). The green
  filled circle indicates the single FSRQ, orange points are the extra
  sources, while the blue triangles denote sources for which we derived an
  estimate of the redshift from the host galaxy contribution to the SED
  \citep[see, e.g.][]{3HSP}. The average value for TXS\,0506+056 
  (magenta point:
  \citealt{Padovani_2019}), which also belongs to our sample, and the
  revised blazar sequence (brown filled squares: \citealt{Ghisellini_2017})
  are also shown. Error bars denote the uncertainties (our sources) and the
  sample dispersion (blazar sequence). The typical \nup~uncertainty is also
  shown (vertical dotted line). Arrows denote lower limits on \nup~(vertical) and
  redshifts (diagonal), which also affect the rest-frame \nup~values.
  Sources above the dashed line are extreme blazars. The small 
  light blue points
  are the control sample of IBLs and HBLs (see Section
  \ref{sec:diff_blazar} for details).}
\label{fig:sequence}
\end{figure}

Fig. \ref{fig:sequence} plots the location of our sources (black circles) 
on the \nup~-- $L_{\gamma}$ diagram (note that by definition our sample 
includes only sources with \nup~$> 10^{14}$ Hz). IBLs cover the $L_{\gamma} 
\sim 10^{42} - 10^{47}$ 
erg s$^{-1}$ range while HBLs occupy the narrower $\sim 10^{44} - 10^{46}$ 
erg s$^{-1}$ region all the way up to \nup $\sim 3 \times 10^{18}$ Hz, although 
there are also some lower limits on redshift, which means higher $L_{\gamma}$ are
very likely. 
Six sources have \nup~$> 1$ keV ($2.4 \times 10^{17}$ Hz) and are therefore 
classified as extreme, with one very close at \nup~$> 1.7 \times 10^{17}$ Hz, given
its redshift lower limit. 

The so-called ``blazar sequence'' \citep{fossati98,ghis98} 
\citep[see also, e.g.][for a detailed discussion and references]{giommibsv1}, 
posits the existence of a strong anti-correlation between (bolometric) luminosity
and \nup, related to the apparent lack of FSRQs of the HBL type. Many masquerading 
BL Lacs fill exactly this void. 
A number of blazar sequence outliers have also been discovered by different groups
(see, e.g. \citealt{Padovani_2019}, for a detailed list and also the very recent 
paper by \citealt{Keenan_2021}). 

\cite{Ghisellini_2017} have revisited the original blazar sequence by using the {\it
Fermi} 3LAC sample \citep{Fermi3LAC}, with the result that \nup~changes quite 
abruptly as a function of $L_{\gamma}$, as shown in Fig. \ref{fig:sequence} 
(brown filled squares). The same figure (originally shown by \citealt{Padovani_2019}) 
indicates also the location in its average state (magenta point) 
of TXS\,0506+056, which is an obvious outlier from the sequence. There are three 
sources quite close to the location of TXS\,0506+056, all of them masquerading, 
plus the FSRQ; two of the extra sources (Section \ref{sec:extra}), both 
masquerading, are also in the same region. The issue of how the properties 
of masquerading sources might differ from those of the other sources is 
discussed in Section \ref{sec:masq_diff}. 

No anti-correlation whatsoever is found between \nup~and $L_{\gamma}$, 
although one could argue that this could be due to the fact that we have a 
\nup~$> 10^{14}$ Hz cut. However, while according to the sequence blazars with
$L_{\gamma} > 10^{45}$ erg s$^{-1}$ should typically have \nup~$< 10^{13}$ Hz,
our sources above this power are spread over more than 4 dex in \nup~between
$10^{14}$ and $\sim 10^{18}$ Hz and occupy a region of parameter space, which is
forbidden by the sequence. This happens also at $L_{\gamma} < 10^{45}$ erg s$^{-1}$,
where blazars should typically have \nup~$\gtrsim 10^{16}$ Hz while we find a similar
spread all the way down to $10^{14}$ Hz, which is the defining limit of our sample. 


\section{The extra sources}\label{sec:extra}

\begin{table*}
 \caption{Extra sample properties.}
 \begin{center}
 \begin{tabular}{lllrrrrrrr}
   \hline
    Name & 4FGL name & ~~~~~$z$& \nup & $P_{\rm 1.4GHz}$ & $L_{\rm [\ion{O}{II}]}$ & $L_{\rm [\ion{O}{III}]}$ & $M_{\rm BH}$ & $L/L_{\rm Edd}$ & $L_{\gamma}/L_{\rm Edd}$ \\
    &  &  & [Hz] & [W Hz$^{-1}$] & [erg s$^{-1}$] & [erg s$^{-1}$] & [$M_{\odot}$] &  &  \\
    \hline
 5BZU J1339$-$2401  & 4FGL J1339.0$-$2400    &  \magg0.655 &  14.3 &  26.75 &  41.6 &  41.8 &  ...~~~  &  -1.5 &  -0.6 \\
 5BZB J1455+0250  &  4FGL J1455.0+0247    & >0.65 &  14.5 & >26.15 &  ...~~~  &  ...~~~  &  ...~~~  &  ...~~~  & >-1.7 \\
 MG3 J225517+2409  & 4FGL J2255.1+2411   & >0.863 &  15.0 & >26.20 &  ...~~~  &  ...~~~  &  ...~~~  &  ...~~~  & >-0.3 \\
 NVSS J232538+164641 & 4FGL J2325.6+1644  &  \magg0.4817 &  15.5 &  25.49 &  41.0 &  40.7 &  ...~~~  &  -1.8 &  -1.7 \\
    
 \hline
\multicolumn{7}{l}\footnotesize{\textit{Notes.} All values, apart from  redshift, are in logarithmic scale.}\\
  \end{tabular}
  \end{center}
 \label{tab:extra_sample}
\end{table*}

Optical spectra of some targets that were included in a preliminary version
of the G20's list were also presented in Paper I. 
These are still blazars 
without a redshift determination, which turned out not to fulfil all the final 
criteria adopted by the latter authors, especially as regards the size of the 
IceCube error ellipse and the \nup~cut. We discuss here the four sources with 
\nup~$> 10^{14}$ Hz with the aim of characterising these sources and looking 
for other masquerading BL Lacs. These sources are not included in the 
statistical analysis of Section \ref{sec:characterisation} (but see below for the 
$L_{\gamma} - P_{\rm 1.4GHz}$ correlation).

\begin{table*}
 \caption{Masquerading BL Lac properties: extra sample.}
 \begin{center}
 \begin{tabular}{lclcccc}
   \hline
    Name & IceCube Name & ~~~~~$z$ & $P_{\rm 1.4GHz} - L_{\rm \ion{O}{II}}$ & $P_{\rm 1.4GHz}$ & $L/L_{\rm Edd}$ & $L_{\gamma}/L_{\rm Edd}$ \\
      \hline
 5BZU J1339$-$2401     & IC131202A &  \magg0.655 & \checkmark &  \checkmark &  \checkmark &  \checkmark \\
 5BZB J1455+0250     & IC111201A & >0.65 & --  & \checkmark & --  & --  \\
 MG3 J225517+2409    & IC100608A & >0.863 & --  & \checkmark & --  & \checkmark \\
 NVSS J232538+164641 & IC140522A &  \magg0.4817 & \checkmark & I  & \checkmark & \xmark  \\
  \hline
  \end{tabular}
  \end{center}
  \footnotesize {\textit{Notes.} ``I'' implies that the condition is not met but this does not mean this is not a masquerading BL Lac, --'' that no information is available.}
 \label{tab:masq_extra}
\end{table*}

Table \ref{tab:extra_sample} gives the main properties of the extra sources, 
where the columns are the same as in Table \ref{tab:sample}. Note that the first
source is just outside the largest error ellipse considered by G20 
(the nominal IceCube major and minor axes scaled by 1.5: Section \ref{sec:detect}),
while the other three are associated with events having an area
of the error ellipse larger than that of a circle with radius r~$=
3^{\circ}$. Both
sources for which we have $L_{\rm [\ion{O}{II}]}$, 5BZU J1339$-$2401 and NVSS 
J232538+164641, are close to the locus of jetted AGN in Fig. 
\ref{fig:Lr_LOII} and are therefore ``bona fide'' masquerading BL Lacs. 
The former object fulfils all other criteria for such sources, as shown in 
Table \ref{tab:masq_extra}, while the latter one has $L/L_{\rm Edd} \sim
0.014$, i.e. it is HEG-like. The other two sources, 5BZB J1455+0250 and MG3 
J225517+2409, were classified as masquerading BL Lacs due to their 
$P_{\rm 1.4GHz} > 10^{26}$ W Hz$^{-1}$ (Fig. \ref{fig:Lr_Lgamma}), with 
the second one having also $L_{\gamma}/L_{\rm Edd} > 0.5$. 
In short, all four extra sources are of the masquerading BL Lac type. 
The extra sources fall on the high $P_{\rm 1.4GHz}$ -- high $L_{\gamma}$ side
of Fig. \ref{fig:Lr_Lgamma}, as is the case for the other masquerading BL Lacs 
(see Section \ref{sec:masq_diff}). The best fit including them is still 
significant at the $> 99.99$ per cent level with $L_{\gamma} \propto 
P_{\rm 1.4GHz}^{0.97\pm0.12}$, fully consistent with our previous fit 
($L_{\gamma} \propto P_{\rm 1.4GHz}^{1.04\pm0.13}$). As for Fig. 
\ref{fig:sequence} the extra sources, again, fall where most masquerading sources
are (Section \ref{sec:masq_diff}). 

MG3 J225517+2409 has been recently reported as a possible counterpart 
of five ANTARES track events in the $\sim 3 - 40$ TeV energy range, on top of  
being within the (large) error region of the IceCube track IC100608A,
with a combined a posteriori ANTARES space-time and IceCube p-value of $2.6\,\sigma$
\citep{Albert_2021}. The fact that this source is also a masquerading BL Lac, like TXS\,0506+056, 
is therefore tantalising. 

\section{Theoretical interpretation}\label{sec:theory}

We present here estimates for the expected high-energy neutrino emission from our 
sample (i.e. 36 IBLs and HBLs with redshift from Paper I, 5BZB J1322+3216
and 3HSP J152835.7+20042, for which we derived a redshift estimate, and 3HSP J095507.9+35510, for a total of 39 sources)
within the context of two simplified theoretical scenarios that are summarised below. We also compute the stacked neutrino flux and compare it 
to the diffuse IceCube neutrino flux.
\begin{itemize}
\item \textit{Scenario A.} High-energy neutrinos are produced by photomeson interactions 
of relativistic protons with jet synchrotron photons of observed energy 
$h\nu_{\rm peak}^{S}/(1+z)$ \citep{Mannheim_1993, Muecke_2003, Petro_2015,Cerruti_2015}. 
The cascade electromagnetic emission, which is induced by photohadronic 
interactions and photon-photon pair production, has a non-negligible contribution in 
the \textit{Fermi}-LAT energy band \citep[see e.g. Figs.~5 and 6 in ][]{Petro_2015}. 
This scenario may also apply to masquerading BL Lacs and FSRQs if the neutrino production 
site lies beyond the BLR \citep[e.g.][]{Gao_2017}, so that jet radiation provides the main 
target photon field; for an application to TXS\,0506+056 see also \cite{Padovani_2019}.
\item \textit{Scenario B}. High-energy neutrinos are produced by photomeson interactions 
of relativistic protons in the jet with the BLR photons of characteristic energy 
(in the AGN rest frame) $\tilde{\varepsilon}_{\rm BLR}$ \citep[e.g.][]{Atoyan_2001,Murase_2014, Rodrigues_2018}. 
This scenario applies to FSRQs and masquerading BL Lacs, assuming that the neutrino 
production site lies within the BLR. 
\end{itemize}

\subsection{Scenario A}
In this scenario photohadronic interactions have an active role in shaping the blazar
SED, as detailed in \cite{Petro_2015}. The low-energy hump of the SED is explained by 
synchrotron radiation of relativistic electrons in the jet (primary electrons). It is 
assumed that protons are accelerated to high enough energies (e.g. multi-PeV to EeV 
energies) as to pion-produce on photons with energy $h\nu_{\rm peak}^{S}/(1+z)$ (in 
the observer's frame). The high-energy hump is then explained by a combination of 
synchrotron-self Compton (SSC) emission from primary electrons and synchrotron 
emission of electrons and positrons (secondary pairs) produced in photomeson 
production and photon-photon pair-production processes (henceforth, $p\pi$ component). 
Since the luminosity of the $p\pi$ cascade component is directly connected to that 
of high-energy neutrinos, this scenario allows us to associate the observed $\gamma$-ray 
luminosity with the expected neutrino luminosity of a blazar. 
Since external radiation fields are not considered in this scenario, 
the true nature of BL Lac objects in our sample will not be a determining factor in the calculations that follow.

In what follows, we adopt the approach of \cite{Padovani_2015b}. The bolometric 
all-flavour neutrino and antineutrino flux of a blazar can be parametrised as 
\begin{equation} 
F_{\nu+\bar{\nu}}= Y_{\nu \gamma} F_{\gamma},
\label{eq:Fnu}
\end{equation}
where $F_{\gamma}=L_{\gamma}/ 4 \pi d_{\rm L}^2$ is the $0.1-100$ GeV $\gamma$-ray flux 
and $d_{\rm L}$ the luminosity distance. 
We caution the reader that the {\it Fermi}/LAT flux alone might not be a good predictor 
of the expected neutrino numbers, as indicated by recent studies of the neutrino excess 
from TXS\,0506+056 \citep[][and references therein]{Reimer_2019,Zhang_2020} and pointed 
out earlier by \cite{Krauss_2018}. The latter's conclusion was based on the premise that 
the integrated neutrino energy flux cannot exceed the integrated electromagnetic energy 
flux between 0.1 keV and 1 TeV. Given the scatter between {\it Fermi}/LAT and 0.1 keV 
– 1 TeV fluxes (see their Fig. 4), differences of about one order of magnitude in predicted 
neutrino numbers are possible between models that rely on the broadband or the {\it Fermi}/LAT 
flux as proxies for the neutrino flux. More complex scenarios are certainly also viable, such as
the production of TeV - PeV neutrinos in coincidence with X-ray flares powered by proton synchrotron 
radiation \citep{Mastichiadis_2021}, which would further enhance the uncertainties on the predicted 
neutrino fluxes. Hence, the results of this scenario should be viewed in light of the above mentioned 
caveat.


The neutrino-to-$\gamma$-ray luminosity ratio, $Y_{\nu \gamma}$, encodes 
information about the baryon loading of the jet and the photomeson production 
efficiency ($f_{\rm p\pi}$) \citep{Murase_2014,Petro_2015,Palladino_2019}. 
Roughly speaking, $Y_{\nu \gamma}\approx (3/8) f_{\rm p\pi} \xi$, where $\xi = L_{\rm p}/L_{\gamma}$ 
is the baryon loading factor and $L_{\rm p}$ 
is the bolometric luminosity in relativistic protons as measured in the observer's 
frame. Because 
$f_{\rm p\pi}$ in this scenario depends strongly on the source conditions (i.e. 
magnetic field strength, Doppler factor, and source radius) we cannot estimate $\xi$ without proper 
modelling of the sources \citep[see also Fig.~15 in][]{Petro_2020}.

The neutrino and antineutrino energy flux (in units of erg cm$^{-2}$ s$^{-1}$ erg$^{-1}$) 
can be approximated by the following expression
\begin{equation}
    F_{\varepsilon_{\nu}} = F_0 \, \varepsilon_\nu^{-s} {\rm e}^{-\varepsilon_{\nu}/\varepsilon_{\nu, 0}}
    \label{eq:flux-scenarioA}
\end{equation}
where $s<0$ and $\varepsilon_{\nu, 0}$ is a characteristic neutrino energy given by
\begin{equation}
\varepsilon_{\nu, 0} \simeq \frac{17.5~{\rm PeV}}{(1+z)^2} \left(\frac{\delta}{10} \right)^2 \left(\frac{\nu_{\rm peak}^{S}}{10^{16}~{\rm Hz}}\right)^{-1}
\label{eq:E_nu}
\end{equation}
with $\delta$ being the Doppler factor of the neutrino emission region. While there is evidence from radio 
monitoring programs that IBL/HBL objects have lower $\delta$ than LBL sources~\citep[e.g.][and references 
therein]{Homan_2021}, these values are unlikely to be representative of the neutrino production site of the jet. 
Variability at $\gamma$-ray energies on timescales of a few minutes and SED modelling of various blazar sub-classes 
yields usually higher Doppler factors than those inferred by radio observations of parsec-scale jets~\citep[e.g.][]{Begelman_2008, Celotti_2008, PD15, Oikonomou_2021}. In what follows, we use $\delta = 10$ for all sources in the sample 
and comment on the impact of a different value on our neutrino predictions. We note 
that the peak energy of the neutrino spectrum in $\varepsilon_{\nu}F_{\varepsilon_{\nu}}$ 
space is $-s/\varepsilon_{\nu, 0}$. We adopt $s=-0.35$, as in \cite{Padovani_2015b}, 
and set 
$Y_{\nu \gamma}=0.13$, which is the most constraining published upper limit so 
far \citep{Aartsen2016}. The normalisation $F_0$ is then derived by substituting 
equation~(\ref{eq:flux-scenarioA}) (after integrating over neutrino energy) into equation~(\ref{eq:Fnu}).
\begin{figure}
        \includegraphics[width=0.47\textwidth]{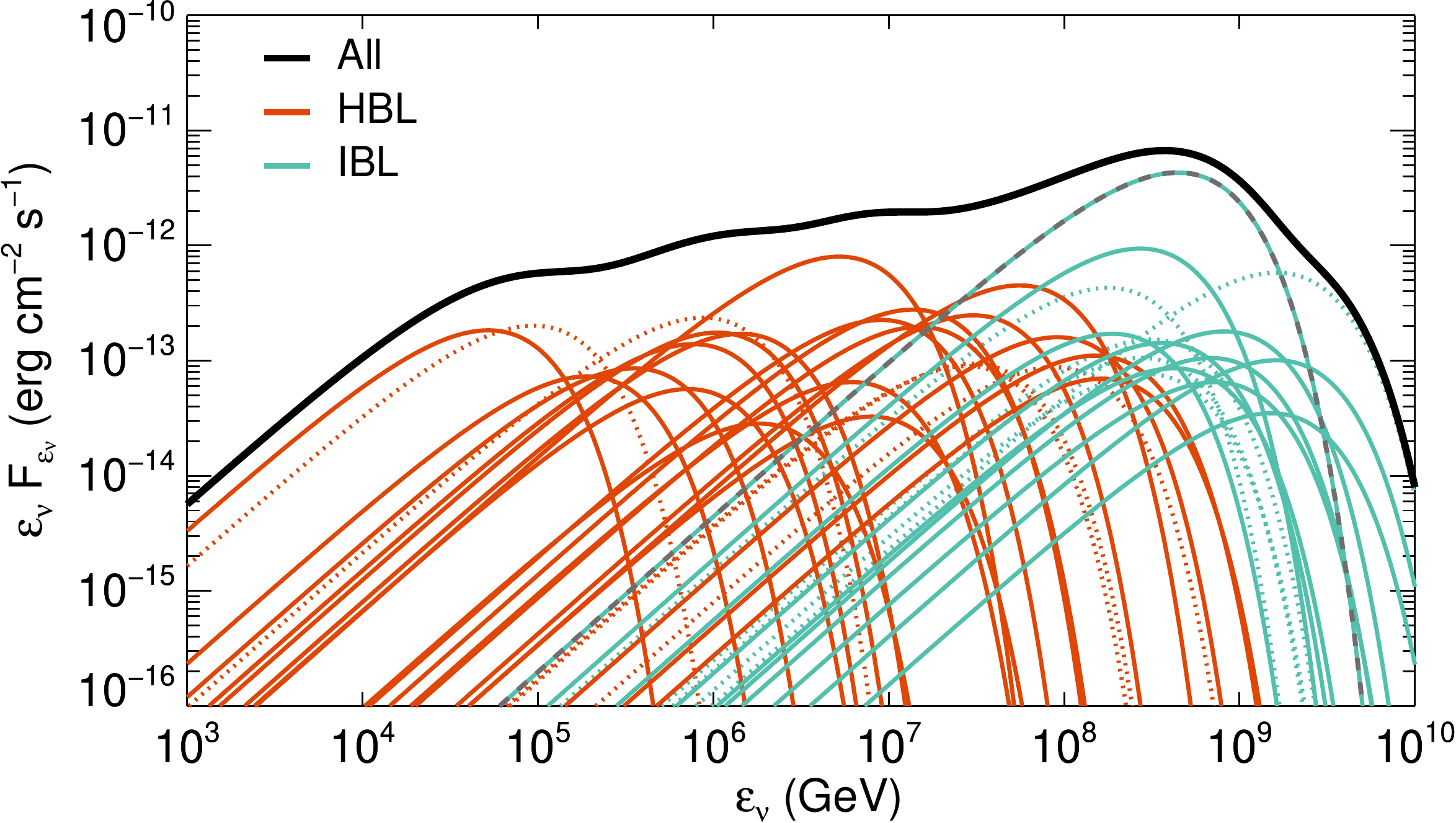}
    
        \includegraphics[width=0.47\textwidth]{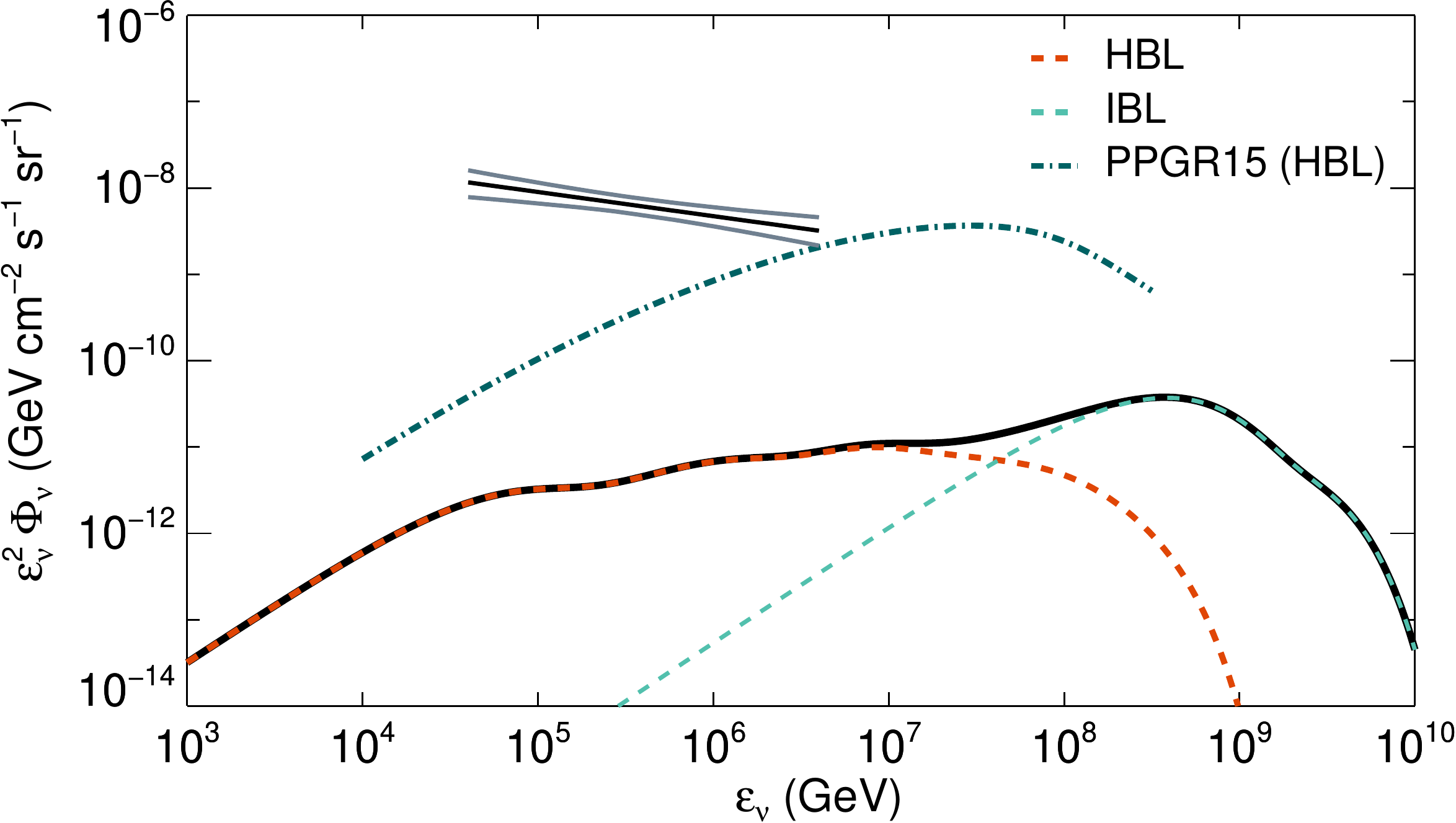}
        
    \caption{\textit{Top panel:} All-flavour neutrino energy flux spectra predicted 
    by Scenario~A for the sources of our sample with measured $z$ and $\nu_{\rm peak}^S$ 
    (thin solid lines). Results for sources with a lower limit on $z$ and $\nu_{\rm peak}^S$ 
    are plotted with thin  dotted lines. The model prediction for 
    TXS\,0506+056 is indicated with a dashed grey line. The total neutrino flux from the 
    sources in the sample is overplotted with a thick black line. A Doppler factor 
    $\delta=10$ and $Y_{\nu \gamma}=0.13$ are assumed for all sources. \textit{Bottom panel:} 
    Stacked single-flavour neutrino flux predicted in Scenario A (see equation~(\ref{eq:stacked-flux})). 
    Dashed coloured lines show the contribution from HBL and IBL objects. The diffuse neutrino 
    flux per flavour from HBL objects for $Y_{\nu \gamma}=0.13$ is also shown (dash-dotted green line; 
    result adapted from \citealt{Padovani_2015b}).
    The updated best-fit astrophysical muon neutrino flux from IceCube, assuming a single power-law 
    in energy, is overplotted  for comparison (solid black line). Solid grey lines indicate 
    the 68 per cent uncertainty range. }
    \label{fig:fluxes_scen_A} 
\end{figure}

Neutrino energy spectra for all sources in the sample with measured $z$ and $\nu_{\rm peak}^S$ are presented in Fig.~\ref{fig:fluxes_scen_A} with solid lines (top panel). Dashed lines are used for sources with lower limits on either $z$ or $\nu_{\rm peak}^S$, for which equation~(\ref{eq:E_nu}) provides an upper limit on $\varepsilon_{\nu, 0}$. Neutrino predictions for HBL and IBL objects are also indicated by different colours. The total neutrino signal from all sources is overplotted with a thick black line. In this scenario, the neutrino spectra of IBL objects peak systematically at higher energies than in HBL. This is a direct result of the dependence of the characteristic neutrino energy on the peak synchrotron frequency (see equation~(\ref{eq:E_nu})). In case of systematic $\delta$ differences between 
the IBL and HBL classes the divide in the neutrino spectra from these two blazar classes would be even more pronounced, 
since $\varepsilon_{\nu,0}\propto\delta^2$. These differences would smear out though, if $\delta$ values were instead randomly distributed around a mean value. The predicted neutrino spectrum of TXS\,0506+056 (dashed grey line) peaks at $\sim 400$~PeV, well above the 90 per cent confidence level upper limit on the neutrino energy of IC~172209A \citep{iconly}. This result is in general agreement with SED models of the 2017 multi-messenger flare, where pion production takes place on jet synchrotron photons \citep[e.g.][]{Cerruti2019}. This should be contrasted to models invoking pion production on more energetic photons of external origin, that may yield peak neutrino energies below 1~PeV \citep[e.g.][]{Keivani2018}. Interestingly, the superposition of individual hard ($s<0$) neutrino spectra results in an almost flat neutrino spectrum in $\varepsilon_\nu F_{\varepsilon_\nu}$ units (thick black line). 

On the bottom panel of the same figure we show the stacked muon neutrino and antineutrino flux over the whole sky defined as
\begin{equation}
    \varepsilon_{\nu}^2\Phi_{\nu_{\mu}+\bar{\nu}_{\mu}}(\varepsilon_{\nu}) = \frac{1}{3}\frac{f_{\rm cor}}{4\pi}\sum_{i=1}^{N=39} \varepsilon_{\nu} F_{\varepsilon_\nu , i},
    \label{eq:stacked-flux}
\end{equation}
where we assumed vacuum neutrino mixing and used 1/3 to convert the all-flavour to muon neutrino flux. Here, $f_{\rm cor}= 16/47$ accounts for the fact that not all sources selected initially by G20 were plausible neutrino emitters. Dashed coloured lines show the contribution of HBL and IBL objects to the stacked signal. Our estimate is compared to the latest IceCube measurements of the astrophysical diffuse neutrino flux adopted from \cite{stettner2019} (black and grey lines). The predicted stacked signal is below the diffuse flux measurement by a factor of $\sim 100$, which can be understood as follows. First, the adopted value for $Y_{\nu \gamma}$ ensures that the diffuse flux from the HBL population is in agreement with IceCube upper limits above 10~PeV \citep{Aartsen2016}. Second, our sample contains only a fraction of the whole HBL population. This becomes clearer if we compare the stacked signal from our HBL sub-sample with the predicted diffuse neutrino flux of the HBL population (dash-dotted green line). This has been computed by \cite{Padovani_2015b} with Monte Carlo simulations of the blazar population, assuming a universal value of $Y_{\nu \gamma}=0.8$, which has been since then constrained to $0.13$~\citep{Aartsen2016}. Hence, the curve displayed in the figure is scaled down by a factor of $0.8/0.13$ compared to the original one in \cite{Padovani_2015b}. The stacked signal from the HBL objects of our sample has a  spectral shape similar to the diffuse spectrum but with the contribution of individual sources being more prominent because of the small size of the sample. 

In summary, if Scenario A is the true physical model for blazar neutrino emission, then the sources identified in our sample are just representative of the whole HBL/IBL blazar population
(see also Section~\ref{sec:detect}), with masquerading BL Lacs not being ``special'' sources in terms of neutrino production. 

\subsection{Scenario B}
\label{sec:ScenarioB} 
In this scenario the photon targets for neutrino production through photomeson interactions with the 
relativistic protons in the jet are provided by the BLR. Hence, this scenario 
is relevant only to the one FSRQ and the 26 sources with an estimate of 
(or an upper limit on) $L_{\rm BLR}$ in our sample, which include all but one 
masquerading BL Lacs. 

The BLR photon field has a luminosity that is a fraction $f_{\rm cov}$ of the disk luminosity $L_{\rm d}$, namely $L_{\rm BLR}= 0.1 \, f_{\rm cov, -1}L_{\rm d}$. We assume that the BLR is a thin spherical shell of characteristic radius $R_{\rm BLR}\simeq 10^{17}\,f_{\rm cov,-1}^{-1/2} L_{\rm BLR, 44}^{1/2}$~cm  \citep{Ghisellini_2008}. Here, we introduced the notation $q_x = q/10^x$ in cgs units, unless otherwise stated. We also assume that the neutrino production region is spherical with comoving radius $R$ and located at distance $r \lesssim R_{\rm BLR} \approx R/\theta_{\rm j}\approx \Gamma \, R$ along a conical jet with bulk Lorentz factor $\Gamma$ and half-opening angle $\theta_{\rm j}$. The accelerated proton distribution is taken to be a power law with slope $p=2$, so that the proton luminosity per logarithmic energy is given by $\varepsilon_{\rm p} L_{\varepsilon_{\rm p}}=L_{\rm p} \mathcal{R}$ with $\mathcal{R}\equiv \ln^{-1}\left(\varepsilon_{\rm p,\max}/\varepsilon_{\rm p,\min}\right)$. 
For simplicity, 
we assume that the proton distribution extends from a minimum energy equal to the proton rest mass energy, $\varepsilon_{\rm p,\min} = \Gamma m_{\rm p} c^2$, to a maximum energy defined by the Hillas confinement criterion \citep{hillas1984}, 
\begin{equation} 
\varepsilon_{\rm p, \max} = \Gamma e B R \approx e B R_{\rm BLR}\simeq 15\, B_{-0.3} \, f_{\rm cov,-1}^{-1/2} L_{\rm BLR, 44}^{1/2}~{\rm EeV},
\label{eq:epmax}
\end{equation}
where $e$ is the proton charge and $B$
is the magnetic field strength in the comoving frame. Photohadronic energy losses lead to similar maximum proton energies, if the acceleration process is fast \citep[see e.g. Table II in][]{Murase_2014}.

The all-flavour neutrino and antineutrino flux (differential in energy) can be written as \citep[e.g.][]{Murase_2014}
\begin{equation}
\varepsilon_{\nu}F_{\varepsilon_\nu}\approx \frac{3}{8}  f_{\rm p\pi} \xi F_{\gamma} \mathcal{R}
\begin{cases}
\left(\frac{\varepsilon_{\nu}}{\varepsilon_{\nu, \rm br}}\right)^2, & \varepsilon_{\nu}\lesssim\varepsilon_{\nu, \rm br} \\
\left(\frac{\varepsilon_{\nu}}{\varepsilon_{\nu, \rm br}}\right)^0, &  \varepsilon_{\nu, \rm br} \lesssim \varepsilon_{\nu} \lesssim 0.05\,\varepsilon_{\rm p, \max}
\end{cases}
\label{eq:flux-scenarioB}
\end{equation}
where $\xi=L_{\rm p}/L_{\gamma}$ is the baryon loading factor and $\varepsilon_{\nu, \rm br} \approx 0.05 \, m_{\rm p} c^2 \epsilon_{\Delta}/(2 \tilde{\varepsilon}_{\rm BLR} (1+z))$ is the typical neutrino energy from pion-production through the $\Delta$ resonance on BLR photons with energy $\tilde{\varepsilon}_{\rm BLR}=10.2$~eV; the latter is measured in the AGN rest frame. Contrary to Scenario A, the photomeson production efficiency is independent of the Doppler factor; it only depends on measurable properties of the BLR, i.e. $f_{\rm p\pi}\approx \tilde{n}_{\rm BLR} \hat{\sigma}_{\rm p\pi} R_{\rm BLR}\approx L_{\rm BLR}\hat{\sigma}_{\rm p\pi}/(4\pi c R_{\rm BLR}\tilde{\varepsilon}_{\rm BLR})$, where $\tilde{n}_{\rm BLR}$ is the number density of BLR photons in the AGN rest frame. The effective cross section is approximated as $\hat{\sigma}_{\rm p\pi}\approx \kappa_{\Delta} \sigma_{\Delta}\delta \epsilon_\Delta/\epsilon_\Delta$, where $\kappa_{\Delta}\simeq 0.2$ and $\sigma_{\Delta}\simeq 5\times10^{-28}$~cm$^2$ are respectively the inelasticity and cross section of the interaction at the $\Delta$ resonance, $\delta\epsilon_\Delta \approx 0.2 $~GeV and $\epsilon_\Delta \approx 0.34$~GeV is the  energy of the $\Delta$ resonance in the proton's rest frame \citep{Dermer_Menon_2009}. 

Meanwhile, secondary pairs injected from charged pion decays and photon-photon annihilation of pionic $\gamma$-rays radiate their energy via synchrotron at energies $\varepsilon_{\gamma}\approx 528~B_{-0.3}(\varepsilon_{\rm p}/10~{\rm PeV})^2 (10/\delta)$~MeV with a luminosity comparable to that of neutrinos, namely 
$\varepsilon_\gamma L_{\varepsilon_\gamma}\large|_{\rm p\pi, syn} \approx (5/6) \varepsilon_{\nu}L_{\varepsilon_\nu} \approx (5/16) f_{\rm p\pi}\varepsilon_{\rm p} L_{\varepsilon_{\rm p}}$ \citep{Murase_2018}. The latter expression can also be written as $\varepsilon_\gamma L_{\varepsilon_\gamma}\large|_{\rm p\pi, syn} \approx (5/16) f_{\rm p\pi} \xi L_{\gamma} \mathcal{R} \approx (5/16) f_{\rm p\pi} \xi \varepsilon_\gamma L_{\varepsilon_\gamma} \left(\mathcal{R}/\mathcal{R}_{\gamma}\right)$, where $\mathcal{R}_{\gamma}\equiv \ln^{-1}\left(300~{\rm GeV}/100~{\rm MeV}\right)$, assuming a $\gamma$-ray spectrum with photon index 2.

\begin{figure}
 \vspace{-1.4cm}
     \includegraphics[width=0.47\textwidth]{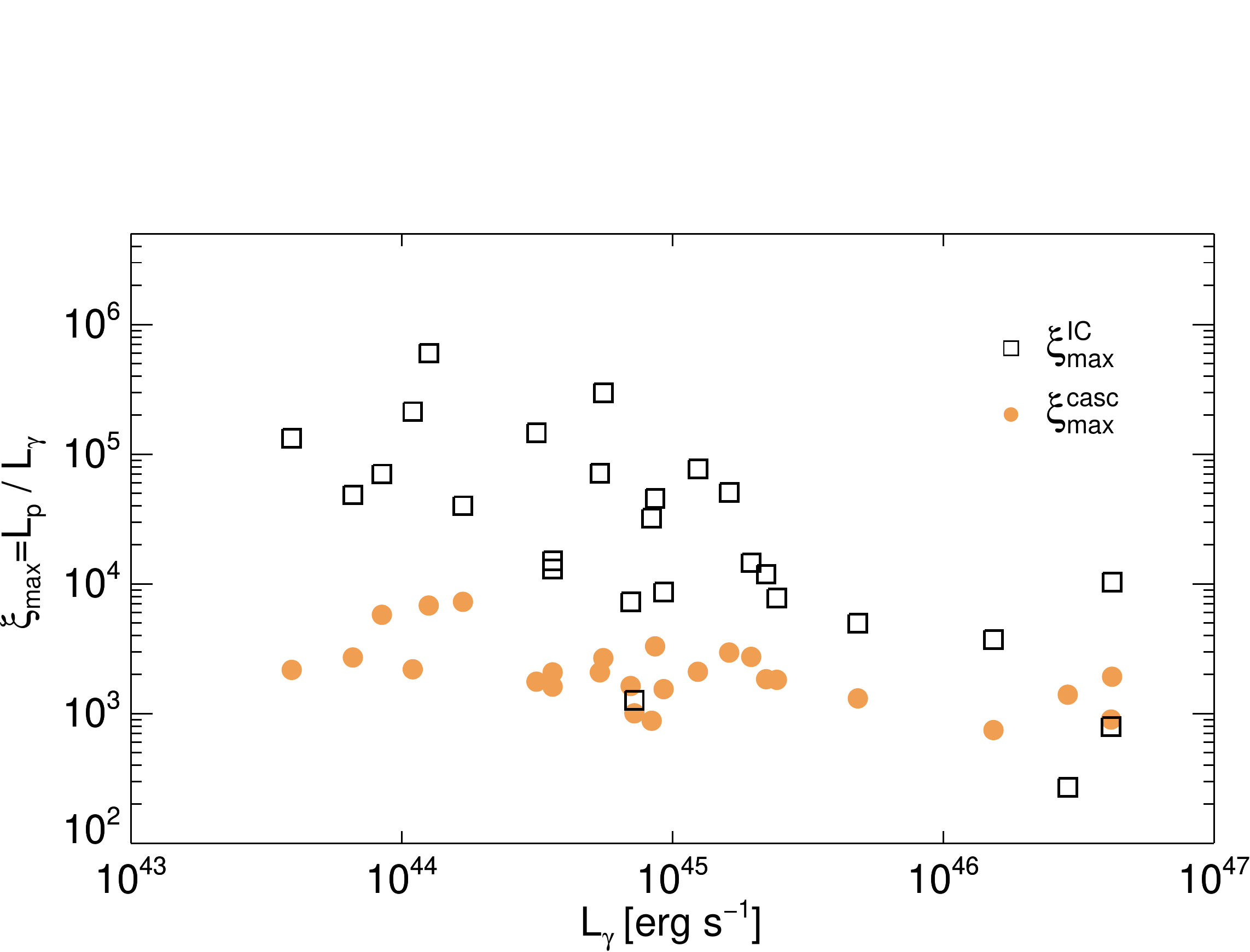}   
    \caption{Maximum baryon loading factor as a function of $L_{\gamma}$
    for the 27 sources with $L_{\rm BLR}$ measurements (or upper limits). Open symbols show the required $\xi$ to match the IceCube sensitivity for each source assuming a $\varepsilon^{-2}_{\nu}$ spectrum. Filled symbols show the limit on $\xi$ imposed by the electromagnetic cascade (see text for details).}
    \label{fig:ximax}
\end{figure}

The baryon loading factor is an uncertain physical parameter \citep[e.g.][]{Murase_2014, Petro_2020}. We therefore compute the value, $\xi^{\rm IC}_{\max}$, needed to match the IceCube sensitivity for each source, namely $(1/3)\varepsilon_{\nu}F_{\varepsilon_\nu} = \varepsilon_{\nu}F_{\varepsilon_\nu}\large|^{\rm IC}_{\rm sens}$.
The declination-dependent sensitivity is a good approximation for the upper limit neutrino flux from the respective sources, because IceCube's point source search has not yet detected any signal. The parametrisation of the sensitivity is taken from \cite{Aartsen2020a}. Another upper limit on $\xi$ can be set from the requirement that the synchrotron cascade luminosity does not overshoot the observed $\gamma$-ray luminosity, namely $\xi^{\rm casc}_{\rm max} = (16/5) f^{-1}_{\rm p\pi} \left(\mathcal{R}/\mathcal{R}_{\gamma}\right)^{-1}$. Figure~\ref{fig:ximax} shows the upper limits on $\xi$ as a function of $L_\gamma$  for 27 sources from our sample with redshift and $L_{\rm BLR}$ measurements (or upper limits) assuming $\delta=10$, $B=0.5$~G and a universal $\varepsilon_{\gamma}^{-2}$ spectrum. The Doppler factor is used only in the estimation of characteristic synchrotron photon energy $\varepsilon_{\gamma}$ and does not enter in the calculation of the neutrino fluxes. Except for two sources, the upper limit on $\xi$ set by the electromagnetic cascade constraint is lower than the value required for matching the IceCube sensitivity (i.e. most filled symbols lie below the open symbols). In other words, the neutrino flux needed to exceed the IceCube sensitivity of most sources would be accompanied by an equally bright cascade emission that would exceed the time-integrated \textit{Fermi}-LAT spectrum. For $\xi = \xi_{\max}^{\rm casc}$ we find $Y_{\nu \gamma}\sim 1$ as expected. 
In addition to the synchrotron cascade component from photopion interactions that we considered here, synchrotron emission from pairs produced in photopair interactions typically emerges in X-rays \citep[for details, see][]{Murase_2018}. For IBL sources, in particular, where the X-ray flux is lower than the $\gamma$-ray flux in the LAT energy range, the constraint set by the photopair cascade emission is expected to be even stronger than the one derived here. 

\begin{figure}
    \includegraphics[width=0.47\textwidth]{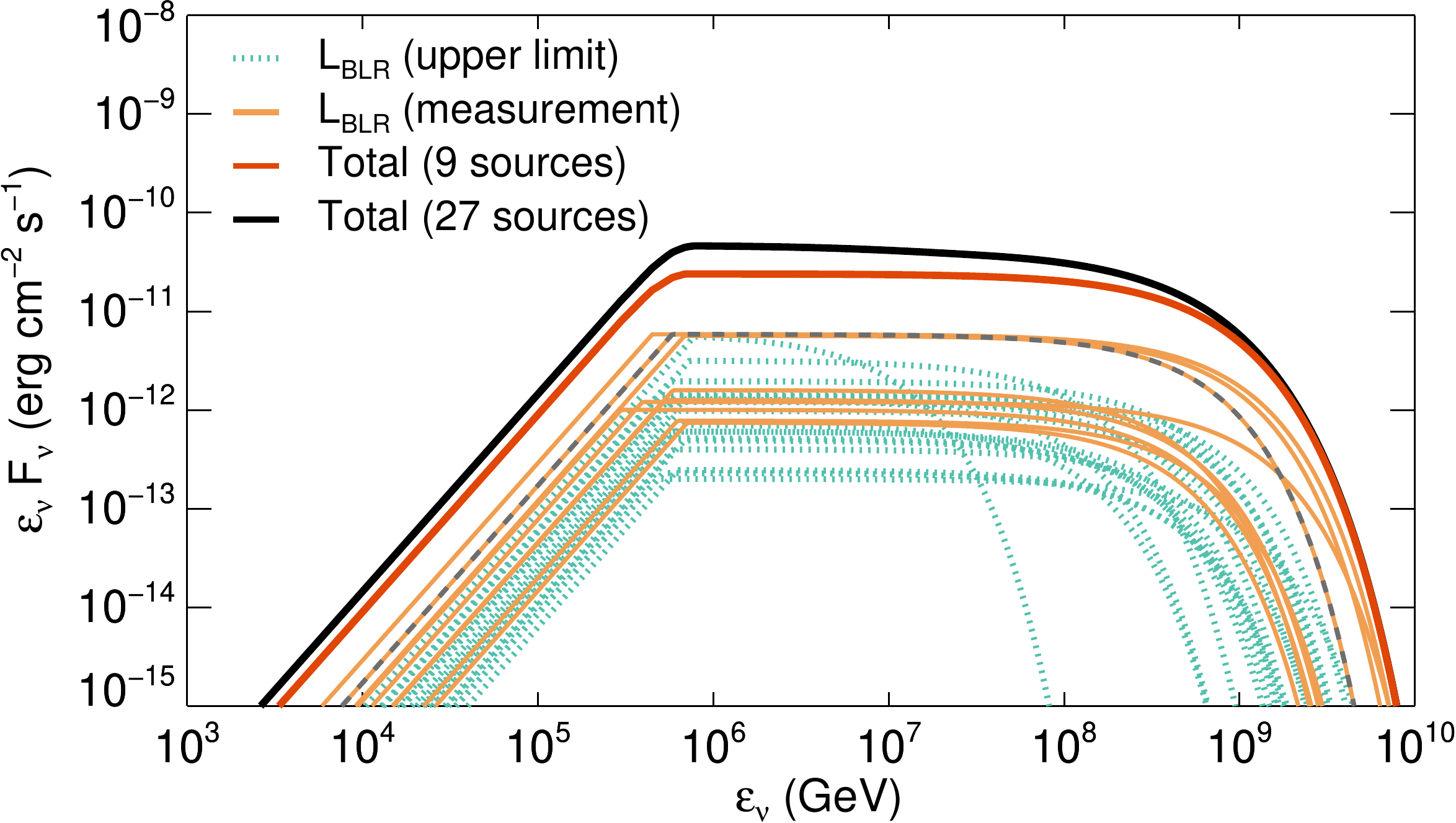}    
    \includegraphics[width=0.47\textwidth]{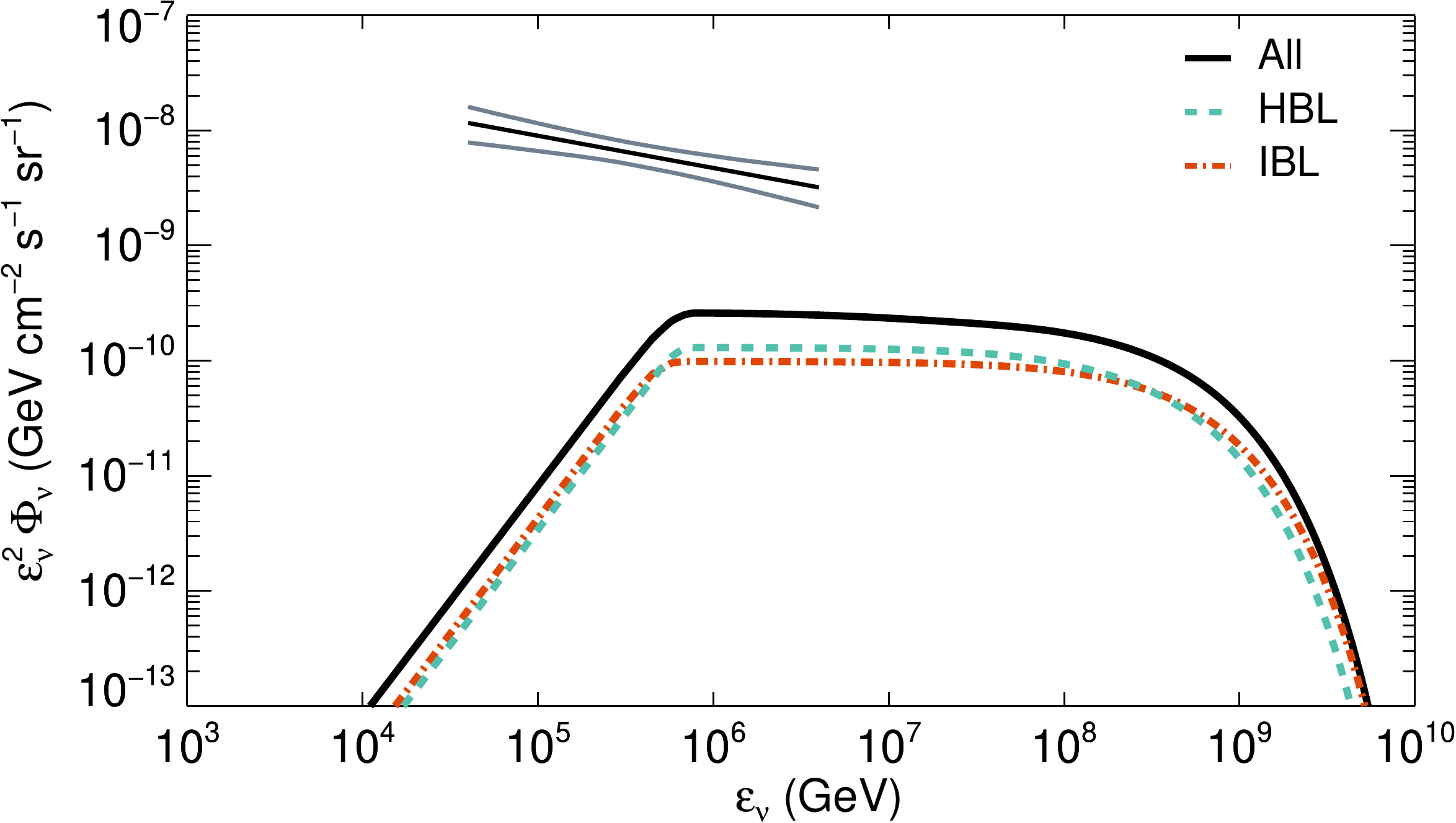}   
    \caption{\textit{Top panel:} All-flavour neutrino energy flux spectra predicted by Scenario~B for the 27 sources with measured $L_{\rm BLR}$ (solid lines) or with upper limits on $L_{\rm BLR}$ (dotted lines). The model prediction for TXS\,0506+056 is indicated with a dashed grey line. The total neutrino flux from all 27 sources is overplotted with a thick black line. The thick solid red line shows the total neutrino spectrum from 9 sources with measured $L_{\rm BLR}$. Other parameters used are: $\tilde{\varepsilon}_{\rm BLR}=10.2$~eV,  $B=0.5$~G, and $\xi=\min(\xi_{\max}^{\rm casc}, \xi_{\max}^{\rm IC})$. \textit{Bottom panel:} Stacked single-flavour neutrino flux predicted in Scenario B using all 27 sources (see equation~(\ref{eq:stacked-flux})). Dashed and dash-dotted coloured lines show the contribution from HBL and IBL objects. The updated IceCube best-fit astrophysical muon neutrino flux, assuming a single power-law in energy, is overplotted  for comparison (solid black line). Dashed grey lines indicate the 68 per cent uncertainty range.}
    \label{fig:fluxes_scen_B}
\end{figure}

Using as baryon loading factor the most constraining upper limit, i.e. $\xi=\min(\xi_{\max}^{\rm casc}, \xi_{\max}^{\rm IC})$, we compute the individual all-flavour neutrino spectra and the stacked muon neutrino flux using equations~(\ref{eq:flux-scenarioB}) and (\ref{eq:stacked-flux}), respectively.
To make the high-energy cutoff of the neutrino spectrum smoother and given that 
the proton spectrum is not expected to have a sharp high-energy cutoff, we multiplied the second branch of equation~(\ref{eq:flux-scenarioB}) with the term $\exp\left(-20 \varepsilon_{\nu}/\varepsilon_{\rm p, max} \right)$. 
Our results are presented in Fig.~\ref{fig:fluxes_scen_B}. In this scenario, the  neutrino spectra from individual sources are similar, with the high-energy cutoff being determined by the maximum proton energy (see equation~(\ref{eq:epmax})). For 18 out of 27 sources of our sample we could derive only upper limits on $L_{\rm BLR}$ (dotted lines in the top panel). However, the 9 sources with measured $L_{\rm BLR}$ (including TXS\,0506+056) make up $\sim57$ per cent of the total energy-integrated neutrino flux of the sample (compare thick black and red lines). Contrary to Scenario A, there is no specific energy range where HBL or IBL objects dominate the stacked signal, since the target photon field is provided by the BLR and is similar in all sources (compare bottom panels in Figs.~\ref{fig:fluxes_scen_A} and \ref{fig:fluxes_scen_B}).  Our stacked single-flavour neutrino signal is lower than the IceCube diffuse flux and extends to EeV energies \citep[see also][]{Murase_2014}. The model prediction is still optimistic, since the baryon loading factor might be limited to even lower values than those shown in Fig.~\ref{fig:ximax} by the photopair cascade emission. In fact, if Scenario B, which implies $Y_{\nu \gamma} =1$, were in operation in all IBL/HBL sources, it would contradict the upper limit derived by IceCube on $Y_{\nu \gamma}$~\citep{Aartsen2016}. Therefore, Scenario B should be considered a generous upper limit
at work only in a small fraction of sources. 

In summary, if Scenario B is the true physical model for blazar neutrino emission, then only a subsample of the HBL/IBL population (with evidence for the presence of external radiation fields) is a neutrino emitter, and no significant differences in the neutrino spectra of HBL and IBL sources are expected.

\section{Discussion}\label{sec:discussion}

\subsection{Are our sources different from the rest of the blazar population?}\label{sec:diff_blazar}

To check if our sources are any different from the rest of the blazar
population we would need a large sample of $\gamma$-ray detected BL Lacs
with \nup~$> 10^{14}$ Hz, i.e. IBLs plus HBLs, which unfortunately does not
exist. \cite{3HSP} have put together the largest sample of HBLs, the 3HSP,
which includes more than 2,000 sources, 88 per cent of which have a
redshift estimation. To make up for the missing IBLs we have compiled a
sample of $\gamma$-ray selected  
objects of this type by accurately verifying \nup~for all the sources 
classified as IBLs (or HBLs and not included in the 3HSP 
sample) in the  {\it Fermi} 4LAC catalogue \citep{4LAC}. 
To obtain a sample that is accurate and has a high level of completeness, we 
have assembled the SED of each candidate located at Galactic latitudes 
$|b_{\rm II}| >30^{\circ}$ combining the multi-frequency information provided 
by the latest version of the VOU-Blazar tool \citep[V1.92, ][]{vou-blazar}, 
which provides access to data from 67 multi-frequency catalogues and spectral 
databases, the data retrieved using the SSDC SED 
tool\footnote{\url{https://tools.ssdc.asi.it/SED/}}, and the results of a 
recent analysis of Swift-XRT data carried out in the framework of the 
Open Universe initiative \citep{giommi2019}. Sources with \nup~between 
$10^{14}$ and $10^{15}$ Hz were selected by visual inspection of the 
available SED data. 
The final IBL sample includes 183 sources, of which 73 have redshift. 
By adding the 600 HBLs with $|b_{\rm II}| > 30^{\circ}$, we obtain
an IBL plus HBL control sample of 783 objects, 630 of which have redshift,
detected at the 100 and 97 per cent level in the $\gamma$-ray and radio
bands, respectively.


The location of these objects in the $L_{\gamma} - P_{\rm 1.4GHz}$ plane is
shown in Fig. \ref{fig:Lr_Lgamma}, where they appear to populate
the same region as our sample. This is confirmed by a variety of
statistical tests: the two samples have similar mean radio and $\gamma$-ray
powers and distributions. Moreover, the control sample displays a very
strong linear correlation between the two powers, significant at the $>
99.99$ per cent level (and not due to the common redshift dependence) with
$L_{\gamma} \propto P_{\rm 1.4GHz}^{0.83\pm0.04}$, consistent with the
correlation found for our sample ($L_{\gamma} \propto P_{\rm
  1.4GHz}^{1.04\pm0.13}$). 

As for the \nup~-- $L_{\gamma}$ plane, Fig. \ref{fig:sequence} shows that the 
IBL plus HBL control sources, as was the case for Fig. \ref{fig:Lr_Lgamma}, 
appear to populate the same region as our sample. Again, this is confirmed 
by a variety of statistical tests: the two samples have similar mean \nup~
and distributions. Here we have included the sources without redshift 
in the IBL sample to increase the statistics (the issue for $L_{\gamma}$ has been
discussed above) and excluded the FSRQ from our sample (M87 is already excluded 
since its \nup~is hard to determine, see Section \ref{sec:sample}). Actually, 
being $\sim 17$ times larger, the control sample shows that many blazars reach 
well into the blazar sequence forbidden zone ($L_{\gamma} > 10^{45}$ erg s$^{-1}$ and 
\nup~$> 10^{13}$ Hz).  

We find a fraction of extreme sources in our HBL sub-sample of $0.25^{+0.18}_{-0.11}$ 
(6/24 or $0.29^{+0.19}_{-0.12}$ [where the uncertainties are derived using Poisson
statistics] for 7/24, where we have included the source which is very close to being 
extreme, as discussed in Section \ref{sec:sequence}), 
to be compared with the 3HSP sample \citep{3HSP} value of $0.11^{+0.01}_{-0.01}$. 
While nominally this is a factor $2.5 - 3$ lower, the two values are still consistent 
within our (large) uncertainties. 

\begin{figure}
\vspace{-2.2cm}
\hspace{-0.6cm}
\includegraphics[width=0.55\textwidth]{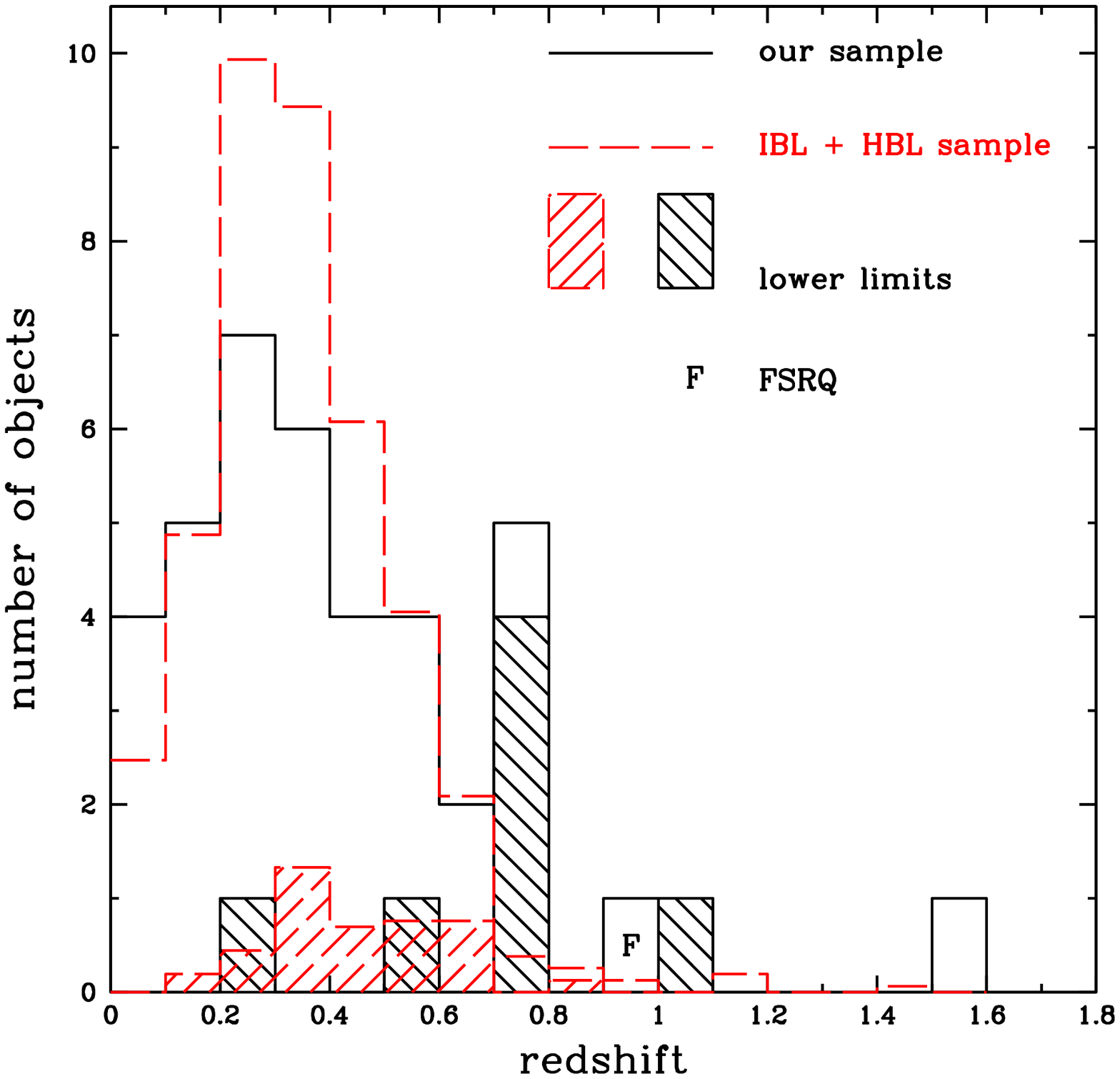}
\vspace{-3.1cm}
\caption{The redshift distribution for our sample (black solid line), with
  lower limits (denoted by the dashed areas); nine sources are still
  without redshift. The FSRQ is
  denoted by an ``F''. The red long-dashed line indicates the control
  sample of IBLs and HBLs 
  scaled to the size of of our sample.}
\label{fig:Nz}
\end{figure}

We also compared the redshift distribution, $N(z)$, of our sources
  (excluding M87 and the FSRQ), to that of the control sample, as shown
  in Fig. \ref{fig:Nz} (black solid and red long-dashed line respectively). 
  To take into account the redshift
  lower limits we used {\tt ASURV} \citep{la92}, the Survival Analysis
  package, which employs the routines described in \cite{fei85} and
  \cite{iso86}, which evaluate mean values by dealing properly with limits
  and also compute the probability that two samples are drawn from the same
  parent population. The mean redshift is $\langle z \rangle =
  0.54\pm0.09$ and $\langle z \rangle = 0.37\pm0.01$ for our sample and 
  the control sample respectively, 
  different at the $\sim 79$ per cent level according to {\tt
    ASURV}\footnote{Here and in the following we use the Peto-Prentice
  test, which, according to the {\tt ASURV} manual, seems to be the least
  affected by differences in the censoring patterns, which are present in
  the two samples.}. The control sample is missing $\sim 20$
  per cent of the redshifts, while we have looked very carefully for even
  weak features and obtained a redshift or a lower limit for all the
  sources we took a spectrum of. Therefore, the fact that the control sample 
  mean redshift is slightly lower than ours is to be expected. This is confirmed
  by comparing the newly determined redshifts in Paper I with the corresponding 
  values in the \cite{3HSP}'s sample (for sources with a previous redshift 
  estimate). Out of seven sources in only one case (3HSP J125848.0$-$04474 a.k.a. 
  4FGL J1258.7$-$0452, for which our redshift is very well determined [see Fig. 1 of Paper I], which means the 
  previous value from the NASA/IPAC Extragalactic Database\footnote{\url{https://ned.ipac.caltech.edu/}} was incorrect) is the 3HSP redshift larger than ours, 
  while for the remaining sources our redshifts are typically higher by 0.1 and up to 
  $\gtrsim 0.2$.
  
\subsection{What is the fraction of masquerading BL Lacs?}\label{sec:fract_masq}

We find nine masquerading BL Lacs (Table \ref{tab:sample}, boldface, and Table 
\ref{tab:masq}) and eight 
sources which fulfil none of our criteria and therefore we consider ``bona fide'' 
non-masquerading objects (Table \ref{tab:sample}, italics). For twenty-one sources, 
however, we do not have the relevant
information to make a decision (nine more sources still have no redshift
information). Therefore, the fraction of masquerading BL Lacs is in the
range $24 - 79$ per cent but should be well above the lower bound given 
our conservative selection (Section \ref{sec:P14_LOII}). Note that 
if we include the four extra sources, all of them masquerading, we obtain
a higher fraction in the $31 - 81$ per cent range. We stress that a 
comparison with other samples is impossible, as nobody has done the 
characterisation we have done here before.

\cite{giommibsv2}, by simulating the blazar population in the {\it Fermi}
2LAC catalogue \citep{Ackermann_2011b}, have found a 70 per cent fraction
of masquerading BL Lacs. 
We note, as they did, that their result is sample and flux limit
dependent. The latest {\it Fermi} catalogues reach fainter $\gamma$-ray
fluxes so one would expect this fraction to decrease, as we reach less
luminous, more LEG-like sources. Our value ($>24$ and $\lesssim 80$ per cent) is,
therefore, consistent with the simulation results. 

A word of caution is in
order: if masquerading BL Lacs turn out to be stronger neutrino emitters than
other blazars, the fraction we derive here is going to be biased on the high side, given
our built-in IceCube selection.

\subsection{Do masquerading BL Lacs have different properties from 
non-masquerading ones?}\label{sec:masq_diff}

All sources in our sample with $L_{\gamma} \gtrsim 10^{46}$ erg s$^{-1}$ in Fig.
\ref{fig:Lr_Lgamma} are masquerading BL Lacs, as are all objects with
$P_{\rm 1.4GHz} \gtrsim 10^{25.5}$ W Hz$^{-1}$ (with the exception of 5BZB
J1322+3216, for reasons discussed in Section \ref{sec:P14_Lgamma}). More 
generally, it appears that
masquerading BL Lacs tend to be more powerful than non-masquerading
ones. This is confirmed for the radio band, where radio powers are $\sim
10$ times larger and for which a Kolmogorov-Smirnov (KS) shows that the two
samples are significantly different ($P \sim 99.8$ per cent). A similar
result ($P \sim 99.2$ per cent) is obtained using {\tt ASURV}. Restricting
the comparison only to the eight ``bona fide'' non-masquerading BL Lacs (excluding 
5BZB J1322+3216, for reasons discussed above) we obtain
practically the same result ($P \sim 98.3$ per cent). It could be argued
that a high $P_{\rm 1.4GHz}$ was one of the criteria for selecting
masquerading BL Lacs but not a single source has been classified as such 
only on the basis of its radio power. $L_{\gamma}$
is also $\sim 6$ times larger for masquerading BL Lacs but a KS test shows
that the two luminosity distributions are not significantly different.
The $P_{\rm 1.4GHz}$ difference is not surprising, as discussed by
\cite{Padovani_2019}, as masquerading BL Lacs need to have relatively 
high powers, besides having \nup $\gtrsim 10^{14}$ Hz, to be able to dilute 
the quasar-like emission lines, and the radio band is closer to the optical/UV 
one than the $\gamma$-ray band. 
Masquerading BL Lacs are also at higher redshift than the rest of the sample  
but not significantly so ($P \sim 15$ per cent): $\langle z \rangle = 
0.61\pm0.19$ vs. $\langle z \rangle = 0.40\pm0.04$, where again we have used 
{\tt ASURV}. This is likely to be related to their higher powers. 

Fig. \ref{fig:sequence} suggests that \nup~might be smaller for 
masquerading BL Lacs. Using {\tt ASURV}, given the two lower limits on
this parameter plus the redshift lower limits, which affects also
the rest-frame \nup, we derive $\langle \log$(\nup)$\rangle = 
15.17\pm0.30$ vs. $\langle \log$(\nup)$\rangle = 16.53\pm0.27$,
with the two distributions being different at the 
$\sim 98.2$ per cent level. 
This difference, again, can be explained by a selection effect: the closer
\nup~is to the band where the lines are diluted, that is the optical--UV
band, the more effective the dilution will be and the more likely the
source will be a masquerading BL Lac. This is borne out by the
$\gamma$-ray simulations mentioned in Section \ref{sec:fract_masq} \citep{giommibsv2}, which
show a steady decrease in the fraction of masquerading BL Lacs as a function
of \nup, going from $\sim 80$ per cent at \nup $\sim 10^{14.5}$ Hz down to
less than $\sim 60$ per cent at \nup $\sim 10^{17.5}$ Hz.

\subsection{Detectability of the theoretically predicted neutrino signal} \label{sec:detect}

We now discuss the detectability of the theoretically predicted neutrino signal in Scenarios A and B with
IceCube and the interpretation of the observed correlation of neutrinos with 
our IBL/HBL sample in the context of these two theoretical scenarios. 

The number of expected muon and anti-muon neutrinos from a single astrophysical point-like source with persistent neutrino emission at
declination $\delta$ in IceCube in the energy range from $\varepsilon_{\nu,{\rm min}}$ to $\varepsilon_{\nu,{\rm max}}$ is,
\begin{equation} 
N_{\nu+\bar{\nu}}~=~\int_{\varepsilon_{\nu,{\rm min}}}^{\varepsilon_{\nu,{\rm max}}} {\rm d} \varepsilon_{\nu} A^{\nu+\bar{\nu}}_{\rm eff}(\varepsilon_{\nu},\delta) \frac{ F_{\varepsilon_{\nu}}}{\varepsilon_{\nu}} \Delta \mathrm{T},
\label{eq:n_nu} 
\end{equation} 
where $A^{\nu+\bar{\nu}}_{\rm eff}(\varepsilon_{\nu},\delta)$ is the effective area of the IceCube detector for neutrinos with energy $\varepsilon_{\nu}$, $F_{\varepsilon_{\nu}}$ is the neutrino flux emitted by each modelled source in Scenarios A and B (given by equations~\ref{eq:flux-scenarioA} and \ref{eq:flux-scenarioB} respectively), and $\Delta \mathrm{T}$ is the time interval under consideration.

We estimate the expected number of signal neutrinos from the sources in our sample using
equation~(\ref{eq:n_nu}). 
We remind the reader that the blazar sample studied here is predominantly based on the correlation study of G20. 
Our goal in this section is to determine whether the neutrino fluxes calculated with Scenarios A and B are consistent with the level of correlation of IBL/HBL blazars with IceCube neutrinos observed by G20. 

The neutrinos which were found to correlate with IBLs and HBLs in G20 are a
heterogeneous sample from several IceCube analyses with different effective areas which evolve with time.
For an estimate, we assume here the effective area of the point-source (PS) analysis
(IC86-2012) of \cite{2017ApJ...835..151A} for all neutrinos except the high-energy starting tracks.
For these we use instead the high-energy starting muon neutrino effective area published
in~\cite{2013Sci...342E...1I}, as it is as much as a factor of $\sim 10$ smaller than the PS effective area at some declinations. The effective areas of other analyses are not publicly available for arbitrary declinations, but can be as much as a factor of four smaller than the PS effective area \citep[see e.g.][]{2017APh....92...30A}. 

For our estimate with equation~(\ref{eq:n_nu}) we assume $\varepsilon_{\nu,{\rm min}} = 100$~TeV, as beyond this energy the number of atmospheric background
neutrinos for declinations $\delta > 0^{\circ}$ can be safely neglected. We have tested our sensitivity calculation by comparing it to the IceCube PS sensitivity 
published in~\cite{Aartsen2020a} for an $\varepsilon_{\nu}^{-2}$ neutrino spectrum. We have very 
good agreement (better than $30$ per cent) in the declination range $0^{\circ} - 35^{\circ}$.
Outside this range our approximate treatment 
differs from the IceCube official one by up to a factor of three. However, only three of the studied sources, which were analysed assuming the PS effective area, lie outside the declination range $0^{\circ} - 35^{\circ}$.

For the 39 sources modelled under Scenario A, and assuming neutrino emission with $100$ per cent duty cycle over the course of $\sim 10$ years spanned by the neutrino sample, i.e. setting $\Delta \mathrm{T} = 10$~yr in equation~(\ref{eq:n_nu}) and assuming $Y_{\nu \gamma} = 0.13$, i.e. equal to the upper limit derived by \cite{Aartsen2016}, we find that the number of muon neutrinos expected in this model is 
\begin{equation}
\sum_{ i = 1 } ^{39}  N_{\nu+\bar{\nu}}( \varepsilon_{\nu} >100~\rm TeV) \sim 0.5.
\label{eq:n_exp_scenarioA} 
\end{equation} 
If all sources from the underlying population of neutrino producing IBLs and HBLs emit neutrinos at a rate proportional to $F_{\gamma}$, we expect the number of neutrinos generated by the IBLs and HBLs found to correlate with the IceCube events by G20 to be 
\begin{equation}
\sum_{i=1}^{39} N_{\nu+\bar{\nu}, i} \sim \frac{\sum^{39}_{i=1} F_{\gamma,i}}{\sum^{N_{\rm tested ~sources}}_{i = 1}F_{\gamma,i}} \times N_{\rm signal ~neutrinos}.
\label{eq:n_exp_scenarioA_population} 
\end{equation} 
Here, 39 is the number of modelled sources, 
$N_{\rm signal ~neutrinos} =  16\pm4$ is the number of signal neutrinos, and 
$N_{\rm tested~sources}$ is the total number of IBLs/HBLs in the tested blazar sample. The latter is the number of IBLs and HBLs in the {\it Fermi-}4FGL-DR2 catalogue
in the area of the sky where the search for matches with IceCube neutrinos 
was performed by G20 ($|{b}| > 10^{\circ}$ and roughly $|\delta| < +35^{\circ}$),
which is 769 (536 HBLs and 233 IBLs). 

Using $N_{\rm tested~sources} = 769$ in equation~(\ref{eq:n_exp_scenarioA_population}) we then obtain $\Sigma_{i=1}^{39}
N_{\nu+\bar{\nu}, i} \sim 0.6\pm 0.2$, consistent with the number of neutrinos predicted by Scenario A for the 39 modelled sources with equation~(\ref{eq:n_exp_scenarioA}). It is interesting to note that the 39 sources in our sample constitute $39/769\sim 5$ per cent of the population by number and $4$ per cent of the total $F_{\gamma}$. Thus they appear to be an unbiased sample of the underlying ISP/HSP distribution in terms of $F_{\gamma}$.

The nine confirmed masquerading BL Lacs in our sample (see Table~\ref{tab:masq}) are responsible for $\sim 60$ per cent of the total $F_{\gamma}$ in the sample of 39. The eight bona-fide non-masquerading sources produce $\sim 10$ per cent of the total flux, while the twenty-one sources which we cannot classify make 
the remaining 30 per cent. However, the nine masquerading BL Lacs produce 25 per cent of the expected neutrinos in IceCube, which is close to 9/39. This is because the neutrino expectation in IceCube depends not only on $F_{\gamma}$ but also on the declination, peak neutrino energy, and the IceCube analysis with which each neutrino is detected.  The neutrino flux fraction carried by the masquerading BL Lacs is of course dependent on our model assumption that the proton luminosity is linearly dependent on the {\it Fermi}-LAT $\gamma$-ray flux of each source. Introducing a different relation between neutrino and electromagnetic flux would affect our estimate as shown, for example, by~\citet{Krauss_2018}. Flares and periods of high activity generally lead to enhanced neutrino production, which scales non-linearly with the $\gamma$-ray flux, see e.g.~\cite{Murase_2014,2015MNRAS.451.1502T,Murase_2018}.

We also apply the procedure outlined above to the 27 sources for which we have a firm estimate or upper limit
on the BLR luminosity assuming Scenario B. In this case, we expect 
$\sum_{ i = 1 } ^{27} N_{\nu+\bar{\nu}}(\varepsilon >100~\rm TeV) = 23.7$ muon neutrinos. 
For this sample the signal is dominated by TXS\,0506+056 which is responsible 
for almost half of the neutrinos expected from the
entire sample.
Considering only the nine sources with a firm BLR luminosity measurement (i.e. excluding the upper limits),
we obtain $\sum_{ i = 1 }^{9} N_{\nu+\bar{\nu}} (\varepsilon_{\nu} 
>100~\rm TeV) = 18.5$. Note that 8/9 of these objects are of the 
masquerading type, which shows, as expected, that these sources make up
most of the signal.

We conclude that if Scenario A is in operation in IBLs/HBLs in general it can account for the observed level of correlation found by G20. If this is the underlying
neutrino emission mechanism, the sources we have identified through correlations are not special, but representative of the underlying neutrino producing
IBLs and HBLs. In the future, Scenario A will be tested by IceCube by means of updated estimates (or upper limits) on the $Y_{\nu\gamma}$ parameter. However, if special conditions are in operation among this subsample of blazars, which are counterparts to IceCube neutrinos in the analysis of G20, in the sense that they do not respect 
the upper limit on $Y_{\nu \gamma}$ derived for the entire population by IceCube, and neutrino emission proceeds at the maximum rate allowed by the upper limit on $\xi$ derived in section~\ref{sec:ScenarioB} 
(i.e., in the limit that all blazar-neutrino correlations found in G20 are due to high-energy neutrino emission from these 27 sources), the entire 
neutrino signal of $16\pm4$ neutrinos observed in G20 can be accounted for.

We note that Scenario B does not take into account possible additional
constraints that can be derived on the maximum neutrino flux from each studied source
based, for example, on the cascade emission from pairs produced in photopair interactions, which typically emerges in the X-ray energy range in IBLs. 
For example, for the flaring SED of TXS\,0506+056 in 2017, this constraint (see equation 13 of \citealt{Murase_2018}) would result in a stronger upper limit on $\xi$ than $\xi_{\rm max}^{\rm casc}$. Similarly, archival X-ray observations of  TXS\,0506+056 could limit $\xi$ to a $\sim10$ times lower value than the one used here \citep[see Fig.~1 of][]{Petro_2020}. In other words, what we have studied here with respect to Scenario B is an idealised limit. If the fraction of masquerading BL Lacs and corresponding BLR luminosities of the entire BL Lac population were known we would be able to obtain stronger limits on Scenario B. Additionally, we are aware that if Scenario B is in operation in masquerading BL Lacs then FSRQs might also be strong neutrino emitters and should exhibit a correlation with IceCube neutrinos, which was not observed in the study of G20 nor in previous papers by some of us. 

Since FSRQs are almost all of the LBL type, this point is linked to the correlations 
with LBL rather than IBL/HBL found by some papers both 
through statistical tests and studies of individual sources.  
The evidence in this case is less robust or still being debated. For example, 
\cite{Kadler_2016} studied a possible association in space and time between 
the blazar PKS\,1424-418, an LBL at $z=1.522$, and a $\sim$ 2 PeV IceCube neutrino. 
The positional uncertainty (15.9$^{\circ}$, 50 per cent radius) of this event, 
however, is  large and the a-posteriori probability of a chance coincidence 
was estimated to be only about 5 per cent. \cite{Hovatta_2021} found associations between radio-flaring 
blazars mostly of the LBL type and 16 IceCube events but only at the $2\,\sigma$
level. \cite{Plavin_2020} cross-correlated a very long baseline 
interferometry (VLBI) flux density-limited AGN sample with ICeCube events with 
energies $> 200$ TeV. They found that AGN positionally associated with IceCube 
events have parsec-scale cores stronger than the rest of the sample at the 
0.2 per cent (post-trial; i.e. $\sim 2.9\,\sigma$) level. The four brightest AGN, which 
they selected as highly probable associations, were 3C 279, NRAO{\bf\,530},  
PKS\,1741$-$038, and OR 103, all blazars of the LBL type. \cite{Plavin_2021} further
correlated the same VLBI sample with all public IceCube data covering seven years
of observations, finding a $3.0\,\sigma$ significance. Combined with their
previous result this leads to a post-trial significance of $4.1\,\sigma$. 
These results are under discussion in the literature \citep{Zhou_2021,Desai_2021}.
Moreover, a comparison between the \cite{Plavin_2020} and the G20's results
is not straightforward because of various factors, including: 1. somewhat different 
IceCube samples; 2. different statistical analysis; 3. different source samples. The 
latter point is worth elaborating upon. On the one hand, most IBLs/HBLs in the 
G20's list have low radio flux densities and therefore cannot belong by construction 
to the VLBI sample used by Plavin et al. On the other hand, strong radio sources are not 
excluded by the selection criteria of G20 and moreover all four LBLs selected
by Plavin et al. are strong {\it Fermi} sources. However, only one of them
(PKS\,1741$-$038, a.k.a. 5BZQ J1743$-$0350) was matched to IceCube events by G20.
NRAO\,530, in fact, is located in the error region of a track that is in the Galactic plane 
($|b_{\rm II}| < 10^{\circ}$), while the other two are outside even the IceCube error 
ellipses scaled by a factor of 1.5\footnote{{G20 have
compensated for the detector systematic uncertainties and related
reconstruction errors by scaling the major and minor axes of the 90 per cent error
ellipses, $\Omega_{90}$, by 1.1, 1.3 and 1.5 times their original size
($\Omega_{90\times1.1}, \Omega_{90\times1.3}, \Omega_{90\times1.5}$
respectively). \cite{Plavin_2020} selected 3C 279 and OR 103 only because their 
minimum pre-trial p-value was obtained by adding a systematic error of 
$0.5^{\circ}$ to the IceCube positions.}}. Finally, 3C 279, despite being the 
brightest radio and $\gamma$-ray source and the only one of these blazars  
included in the list of objects used for the search point sources in the 10-year 
IceCube data set, is not listed among the objects considered as likely neutrino 
emitters \citep{Aartsen2020a}.

We will address the topic of possible differences in the neutrino emission of FSRQs and masquerading BL Lacs in a forthcoming publication. 

 In conclusion, Scenario A roughly corresponds to having a neutrino signal from the entire IBL and HBL population, whereas Scenario B corresponds to having a sub-sample of sources which are individually very bright neutrino point sources. At the moment, both are viable to explain the correlation signal observed by G20 and we cannot definitively distinguish between the two.
 If Scenario A or another mechanism with similar neutrino emission is in operation in IBLs and HBLs, we expect IceCube and other neutrino telescopes to detect neutrinos in the $>$ PeV energy range from the direction of these
 sources in the future. If, on the contrary, no significant flux of $>$ PeV neutrinos is detected, and a much stronger limit on $Y_{\nu\gamma}$ is obtained with future analyses sensitive in the $>$ PeV energy range, Scenario A will start to be constrained. In other words, if the true value of $Y_{\nu\gamma}$ is $\ll 0.13$, Scenario A would not produce sufficient neutrinos to explain the observed neutrino-IBL/HBL correlation. If the correlation signal grows with future datasets, but the limit on $Y_{\nu\gamma}$ becomes stronger, it would suggest that a fraction of the IBLs and HBLs (for example the masquerading BL Lacs) produce neutrinos at a rate larger than the rest of the population.

\section{Conclusions}\label{sec:conclusions}

G20 had selected a unique sample of IBLs and HBLs, $34\pm9$ per 
cent of which should be associated with individual IceCube  
tracks. In Paper I we have presented
optical spectroscopy of a large fraction of these sources, which, in
combination with already published spectra, has allowed us to determine
the redshift, or at least a lower limit to it, for 36 $\gamma$-ray 
emitting blazars. To these we have added two more sources for which the
redshift estimate was photometric and M87 and 3HSP J095507.9+35510, 
the latter an extreme blazar associated with an IceCube track, which was 
announced after the G20 paper was completed.
Four other extra $\gamma$-ray emitting blazars from a preliminary 
version of G20's list were also included. 

Here we have carefully characterised these sources to determine their 
real nature, extending the work done by \cite{Padovani_2019} for 
TXS\,0506+056 to a much larger sample of possible neutrino sources. Our main
goals were to: (1) quantify the presence of masquerading BL Lacs, i.e., FSRQs
in disguise whose emission lines are swamped by a very strong jet, as is the case
for TXS\,0506+056; (2) check if these sources were any different from the rest 
of the blazar population; (3) perform a preliminary theoretical analysis of our sample. 
To this aim we have assembled a set of ancillary data and
have also measured and estimated, in many cases for the first time, 
[\ion{O}{II}] 3727 \AA~and [\ion{O}{III}] 5007 \AA,~and $M_{\rm BH}$, respectively. 

Source characterisation was based on $L_{\rm [\ion{O}{II}]}$ and
$L_{\rm [\ion{O}{III}]}$, $P_{\rm 1.4 GHz}$, $L/L_{\rm Edd}$, and 
$L_{\gamma}/L_{\rm Edd}$. Also thanks to the very high signal-to-noise ratio
of our optical spectra we were able to carry out the {\it first} systematic 
study of masquerading BL Lacs. 

Our main conclusions are as follows:

\begin{enumerate}
    \item we do not find significant evidence for any systematic difference between 
    the sources studied in this paper and 
    other blazars of the same type, i.e. IBLs and HBLs, in terms of their radio
    and $\gamma$-ray powers, \nup, and redshift. This result is based
    on a large control sample that we selected ex novo. Given that only less than 
    half of our sample is expected to be associated with IceCube tracks, it might also 
    be that the lack of such difference is due to our relatively small sample;
    \item the fraction of masquerading BL Lacs in our sample is $>$ 24 per cent and 
    possibly as high as 80 per cent. Although nobody else has derived
    this value before, this fraction is consistent with the simulation
    of a $\gamma$-ray selected blazar sample carried out by \cite{giommibsv2} 
    and also with a scenario where all the signal found by 
    G20 ($\sim 34$ per cent) comes exclusively from masquerading BL Lacs; 
    \item masquerading BL Lacs turn out to be exactly as expected: more 
    powerful than the rest in the radio and $\gamma$-ray band, with a slightly
    smaller \nup. These are the properties, which allow them to effectively
    dilute their strong, FSRQ-like emission lines. Moreover, many of them
    are outliers of the so-called blazar sequence, in the sense that
    they have too large of a \nup~for their $\gamma$-ray power;
    \item MG3 J225517+2409, one of our extra sources, reported to be
    (weakly) associated with ANTARES and IceCube neutrinos, is also a 
    masquerading BL Lac;
    \item  we have estimated upper limits on the neutrino emission under two hypotheses about the 
    target photons for the high-energy protons: (1) photons from the jet (Scenario A); (2) photons from 
    the BLR (when the BLR power could be estimated; Scenario B). At present both scenarios 
    are consistent with the observed level of the blazar-neutrino correlation derived by G20. 
    However, in Scenario A our sample is representative of the underlying IBL/HBL 
    population, whereas in Scenario B the sources investigated are stronger neutrino emitters 
    than the rest and might account for the observed 
    blazar-neutrino correlation signal alone. Both scenarios are testable in the near term with IceCube and future neutrino telescopes, placing a more robust upper limit on the normalisation of Scenario A, while we plan to test Scenario B through a detailed study and systematic modelling of the individual photon-neutrino SEDs, which will be the next phase of this  project.

\end{enumerate}

\section*{Acknowledgments}
This work is based on observations collected at the European 
Southern Observatory under ESO programme 0104.B-0032(A) and 
Gran Telescopio Canarias under the programme GTC24-19B.
We acknowledge the use of data and software facilities from the SSDC,
managed by the Italian Space Agency, and the United Nations ``Open
Universe'' initiative.   The comments of the anonymous referee prompted 
us to clarify some of our statements and led to an improved paper. 
This research made use of IDL colour-blind-friendly colour tables \citep{2017zndo....840393W}.
This research has made use of the NASA/IPAC 
Extragalactic Database (NED), which is operated by the Jet Propulsion 
Laboratory, California Institute of Technology, under contract with the 
National Aeronautics and Space Administration. 
This work is supported by the Deutsche
Forschungsgemeinschaft through grant SFB\,1258 ``Neutrinos and Dark Matter
in Astro- and Particle Physics''. MP acknowledges support from the MERAC Foundation through the project THRILL.
Funding for the Sloan Digital Sky Survey
IV has been provided by the Alfred P. Sloan Foundation, the U.S.
Department of Energy Office of Science, and the Participating Institutions.
SDSS-IV acknowledges support and resources from the Center for High
Performance Computing at the University of Utah. The SDSS website is
www.sdss.org.  SDSS-IV is managed by the Astrophysical Research Consortium
for the Participating Institutions of the SDSS Collaboration including the
Brazilian Participation Group, the Carnegie Institution for Science,
Carnegie Mellon University, Center for Astrophysics | Harvard \&
Smithsonian, the Chilean Participation Group, the French Participation
Group, Instituto de Astrof\'isica de Canarias, The Johns Hopkins
University, Kavli Institute for the Physics and Mathematics of the Universe
(IPMU)/University of Tokyo, the Korean Participation Group, Lawrence
Berkeley National Laboratory, Leibniz Institut f\"ur Astrophysik Potsdam
(AIP), Max-Planck-Institut f\"ur Astronomie (MPIA Heidelberg),
Max-Planck-Institut f\"ur Astrophysik (MPA Garching), Max-Planck-Institut
f\"ur Extraterrestrische Physik (MPE), National Astronomical Observatories
of China, New Mexico State University, New York University, University of
Notre Dame, Observat\'orio Nacional/MCTI, The Ohio State University,
Pennsylvania State University, Shanghai Astronomical Observatory, United
Kingdom Participation Group, Universidad Nacional Aut\'onoma de M\'exico,
University of Arizona, University of Colorado Boulder, University of
Oxford, University of Portsmouth, University of Utah, University of
Virginia, University of Washington, University of Wisconsin, Vanderbilt
University, and Yale University.

\section*{Data Availability}
The flux-calibrated and dereddened spectra are available in the online database ZBLLAC.

\appendix
\section{Host galaxies and black hole mass}\label{sec:Appendix}

\begin{figure*}
    \centering
    \includegraphics[ width=0.95\textwidth]{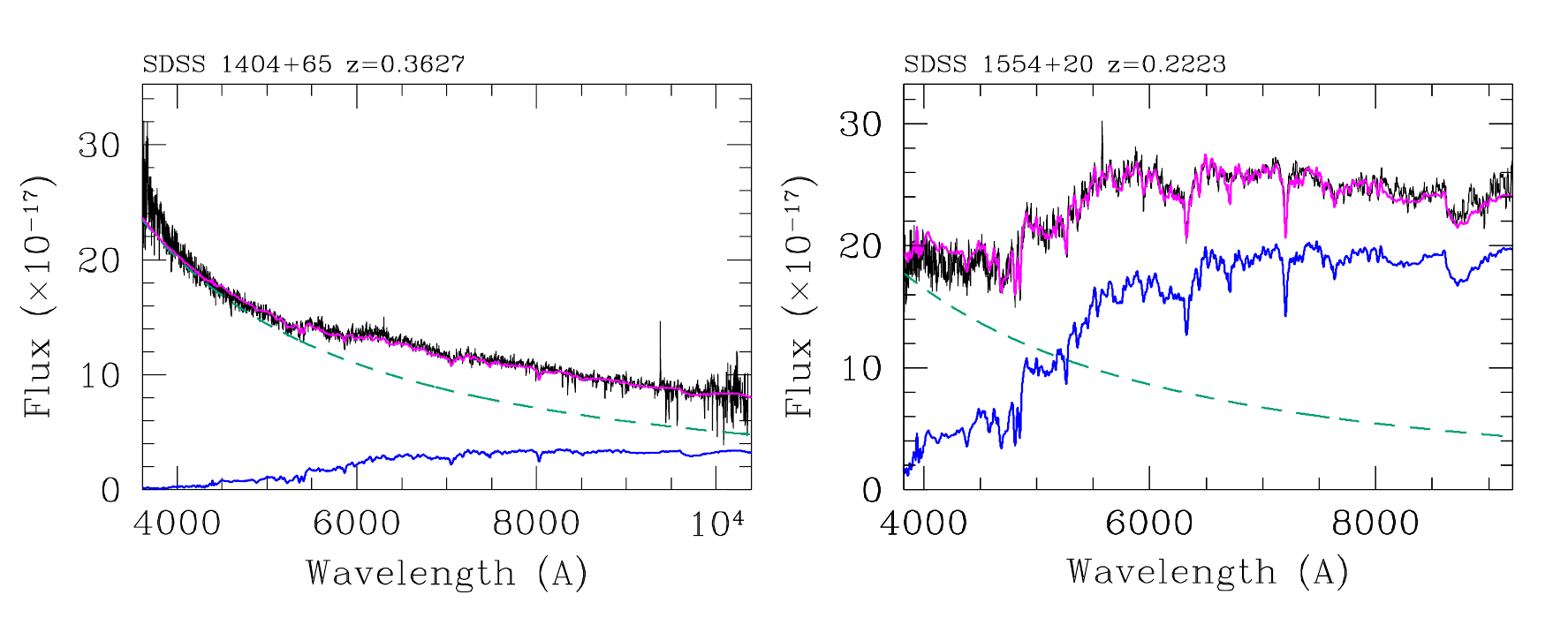}
    \caption{Examples of spectral decomposition for two targets 
    using SDSS spectra. The observed optical spectrum (black line) 
    is fitted by the combination of a power law (green dashed line) 
    plus a template spectrum (blue line) of an elliptical galaxy 
    (see text). The best fit is given by the solid magenta line.
    Fluxes are in units of erg cm$^{-2}$ s$^{-1}$ \AA$^{-1}$.}
    \label{fig:apdx}
\end{figure*}

For eighteen objects with known redshift and a signature of the host galaxy 
in their optical spectra we performed a decomposition of the optical spectra 
as the sum of a power law and an elliptical galaxy template \citep{mannucci2001} 
following the method described in Section \ref{sec:masses} and Paper I.  
Our results are given in Table \ref{tab:apdx} while 
some examples of the spectral decomposition using SDSS spectra 
are given in Figure \ref{fig:apdx}; for  other cases see \cite{Paiano_2021}.
In five further cases the spectrum was not available in electronic format 
or was not flux calibrated. Therefore, we measured the host galaxy magnitude 
directly from an $r$ band image.  

$M_{\rm BH}$ was then evaluated. The dispersion on the
$M_{\rm BH} - M(R)$ relationship is $\sim 0.45$ dex \citep{labita07} and 
we combined this uncertainty with the error on the host galaxy magnitude. 
For thirteen of these objects \cite{Paliya_2021} have derived independently 
$M_{\rm BH}$ based on measurements of the stellar velocity dispersion, $\sigma$, 
and the $M_{\rm BH} - \sigma$ relationship \citep{Gultekin2009} 
that was first applied to BL Lacs by \cite{Falomo_2002} and \cite{Barth2002}. 
For ten of them the two $M_{\rm BH}$ estimates are in 
reasonable agreement, with a difference $\lesssim$ 0.5 dex. 
For three sources, however, (3HSP 023248.5+20171, 3HSP J094620.2+01045, 
3HSP J180849.7+35204), a significant difference 
($>$ 1 dex) was found. The case of 3HSP J094620.2+01045 is explained by 
the different redshift (0.128 vs. 0.576) used by \cite{Paliya_2021}; the 
latter value is the SDSS one and is confirmed by a careful inspection of the spectrum.
 In the case of 3HSP 023248.5+20171 \cite{Paliya_2021} report log$(M_{\rm BH}/M_{\odot}) = 10.8$ 
 based on the measurement of $\sigma$ from a paper copy of the spectrum 
 published by \cite{Perlman1996} that is dominated by the contribution of the 
 host galaxy. The SDSS magnitude of  this object is r = 16.2. 
 Assuming that the whole flux is from the host galaxy the absolute 
 magnitude is $M(R) = -22.9$ that implies log($M_{\rm BH}/M_{\odot}) = 8.8$. 
 Finally, for 3HSP J180849.7+35204 again \cite{Paliya_2021} 
 used a spectrum obtained by \cite{PenaHerazo2019} to obtain 
 log($M_{\rm BH}/M_{\odot}) = 9.36$ based on $\sigma$, while using the 
 same spectrum (available 
 in ZBLLAC) and our spectral decomposition we obtained log($M_{\rm BH}/M_{\odot}) = 
 8.0$. This difference is inconsistent with the dispersion of 
the two $M_{\rm BH} - \sigma$ and $M_{\rm BH} - M(R)$ relationships.
 
\begin{table*}
\caption{Measurement of host galaxy luminosity and black hole mass }
\begin{tabular}{lllllll}
\hline 
Name & $z$ & $M(R)$  & N/H & log$(M_{\rm BH}/M_{\odot}$) & Method & Reference \\
\hline 
3HSP J010326.0+15262 & 0.246 & $-$22.4 &  0.5 &   $8.6\pm0.4$ &   SpD &      SDSS \\ 
5BZU J0158+0101 & 0.454 & $-$20.0 &  5.0 &   $7.4\pm0.6$ &   SpD &    ZBLLAC \\ 
3HSP 023248.5+20171 & 0.139 & $-$22.9 &  0.0 &   $8.8 \pm   0.4$ &   ImD &      SDSS \\ 
3HSP J033913.7$-$17360 & 0.066 & $-$22.5 &  0.0 &   $8.6 \pm   0.4$ &   ImD & PanSTARRS \\ 
CRATES J052526$-$201054 & 0.091 & $-$21.5 &  0.3 &   $8.1 \pm   0.4$ &   SpD &    ZBLLAC \\ 
3HSP J062753.3$-$15195 & 0.31 & $-$23.2 &  0.5 &   $9.0 \pm   0.4$ &   SpD &    ZBLLAC \\ 
3HSP J085410.1+27542 & 0.493 & $-$23.2 &  1.8 &   $9.0 \pm   0.4$ &   SpD &    ZBLLAC \\ 
3HSP J094620.2+01045 & 0.576 & $-$22.5 &  5.0 &   $8.6 \pm   0.6$ &   SpD &      SDSS \\ 
3HSP J095507.9+355101 & 0.557 & $-$22.9 &  0.0 &   $8.8 \pm   0.4$ &   ImD &       P20 \\ 
3HSP J111706.2+20140 & 0.138 & $-$21.4 &  2.5 &   $8.1 \pm   0.4$ &   SpD &      SDSS \\ 
3HSP J123123.1+14212 & 0.256 & $-$22.0 &  2.5 &   $8.4 \pm   0.4$ &   SpD &      SDSS \\ 
3HSP J125821.5+21235 & 0.627 & $-$21.2 &  5.0 &   $8.0 \pm   0.6$ &   SpD &    ZBLLAC \\ 
3HSP J125848.0$-$04474 & 0.418 & $-$22.1 &  4.5 &   $8.4 \pm   0.5$ &   SpD &    ZBLLAC \\ 
5BZB J1322+3216 & 0.4 & $-$21.7 &  4.0 &   $8.2 \pm   0.5$ &   SpD &      SDSS \\ 
VOU J135921$-$115043 & 0.242 & $-$22.3 &  0.5 &   $8.6 \pm   0.4$ &   SpD &    ZBLLAC \\ 
3HSP J140449.6+65543 & 0.362 & $-$21.8 &  5.0 &   $8.3 \pm   0.6$ &   SpD &      SDSS \\ 
VOU J143934$-$252458 & 0.16 & $-$22.3 &  0.0 &   $8.5 \pm   0.4$ &   ImD & PanSTARRS \\ 
3HSP J143959.4$-$23414 & 0.309 & $-$22.3 &  2.5 &   $8.6 \pm   0.4$ &   SpD &    ZBLLAC \\ 
3HSP J144656.8$-$26565 & 0.331 & $-$22.4 &  1.5 &   $8.6 \pm   0.4$ &   SpD &    ZBLLAC \\ 
3HSP J153311.2+18542 & 0.307 & $-$22.1 &  2.0 &   $8.4 \pm   0.4$ &   SpD &      SDSS \\ 
3HSP J155424.1+20112 & 0.222 & $-$22.5 &  0.5 &   $8.6 \pm   0.4$ &   SpD &      SDSS \\ 
3HSP J180849.7+35204 & 0.141 & $-$21.3 &  1.5 &   $8.0 \pm   0.4$ &   SpD &    ZBLLAC \\ 
3HSP J213314.3+25285 & 0.294 & $-$22.8 &  0.0 &   $8.8 \pm   0.4$ &   ImD & PanSTARRS \\ 
 
\hline
\label{tab:apdx}
\end{tabular}
\raggedright\\
\footnotesize{\textit{Notes.} Col. 1: name; Col. 2: redshift; Col. 3: absolute magnitude of the host galaxy 
in the R filter; Col. 4: N/H, i.e. the ratio of the fluxes of the nucleus 
to the host galaxy at 6000 \AA; Col. 5:  the logaritmih of $M_{\rm BH}$ in 
solar units (see text); Col. 6: Method: SpD = Spectrum decomposition; ImD = 
Image decomposition; Col. 7: reference to the spectrum or the image: 
P20 \citep{paiano2020}, SDSS \citep{Ahumada2020}, ZBLLAC \citep{landoni2020}, 
PanSTARRS (\url{https://panstarrs.stsci.edu/})}. \\

\end{table*}

\subsection{Notes for individual targets}
{\bf 3HSP J095507.9+355101}

The measurement of log($M_{\rm BH}/M_{\odot}) = 8.8$ 
for this source is based on the absolute magnitude of the host galaxy 
from the R band image by \cite{paiano2020}. This value is different from the one
previously reported by these authors as that was based on different cosmological parameters. 



{\bf 5BZB J1322+32}

We fitted the observed SDSS optical spectrum into a power law (dominant 
in the blue region) and a galaxy template. An acceptable fit was obtained 
for $z=0.4$ and a galaxy of $M(R) = -21.7$. From this decomposition 
log($M_{\rm BH}/M_{\odot}) = 8.2$ is derived. Note that a redshift value 
$z=0.396$ is reported in \cite{Paliya_2021} based on the  measurement of 
the stellar velocity dispersion from the SDSS spectrum, which is however 
featureless and which reports a value 
derived from the automatic SDSS procedure $z = 1.4\pm0.8$.

\label{lastpage}

\bsp	

\end{document}